\documentclass{llncs}
\usepackage{llncsdoc}
\usepackage{amsmath, amssymb}
\usepackage{amscd,amsfonts,amssymb, color}
\usepackage{latexsym, graphicx, pstricks}
\newtheorem{prop}{Proposition}
\newtheorem{defin}{Definition}
\newtheorem{rem}{Remark}
\newtheorem{lem}{Lemma}
\long\def\symbolfootnote[#1]#2{\begingroup%
\def\thefootnote{\fnsymbol{footnote}}\footnote[#1]{#2}\endgroup}

\newcommand{\sa}{crossover~}
\newcommand{\sae}{crossover~}
\newcommand{\sg}{\Sigma}

\newcommand{\ve}{\varepsilon}
\newcommand{\h}{$H$}
\newcommand{\cd}{\cdot}
\newcommand{\p}{\mathrm{Prefix}}
\newcommand{\s}{\mathrm{Suffix}}
\newcommand{\sub}{\mathrm{sub}}
\newcommand{\rs}{\hspace{-3.5mm}-\hspace{-2mm}<}
\newcommand{\ers}{>\hspace{-2mm}-\hspace{-2mm}<}
\begin{document}
\title{ Generalised sequential crossover of words and languages}
\author{L Jeganathan \and R Rama \and Ritabrata Sengupta}
\institute{Department of Mathematics\\ Indian Institute of Technology
\\Chennai 600 036, India\\ {\tt lj, ramar, rits@iitm.ac.in}}
\maketitle
\begin{abstract}
In this paper, we propose a new operation, Generalised Sequential
Crossover (GSCO) of words, which in some sense an abstract model of
crossing over of the chromosomes  in the living organisms. We extend
GSCO over language $L$ iteratively ($GSCO^*(L)$ as well as iterated
GSCO over two languages $GSCO^*(L_1,L_2)$). Our study reveals that
$GSCO^*(L)$ is subclass of regular languages for any $L$. We compare
the different classes of GSCO languages with the prominent
sub-regular classes. 
\end{abstract}
\section{Introduction}

\par Self-assembly is a process in which smaller objects selectively
aggregate with each other into a complex structure, which in turn
self assemble into larger aggregates. It is a process wide spread in
nature - atoms self assemble into molecules, molecules into
crystals, cells into
 tissues, etc. It is an important tool  in nano-technology, since it takes
 nature as a model and tries to assemble structures from the atomic level
 (bottom-up approach). Self-assembly is considered as a promising technique
 in nano-technology, enabling the fabrication of small complex objects - such
 as computer circuits.

\par A particular case of self assembly is that of a linear self assembly,
 in which one dimensional objects such as DNA double strands interact with
  each other to form  longer strands. DNA recombination is  one such DNA
   self assembly by which Adleman solved an instance of Hamiltonian path
   problem \cite{Ad}. For more than a decade now, self assembly is the
   core of most experiments in DNA computing starting with the celebrated
   experiment of Adleman \cite{Ad,FCL,OK}. Recent developments in DNA computing
    have highlighted the intimate connection between self assembly and computation.
     Computational utilities of DNA self  assembly is studied in \cite{Win}.

\par Most complexity    theoretic studies of self assembly utilise mathematical models.
  Some alternate models, like self assembly of the objects by the use
  of capillary force, electrostatic force, and magnetic force were also studied.

\par In recent years, one can see convergent interests  in the study of
self assembly from Mathematics, Computer science, Physics,
Chemistry, and Biology point of view.
 Yet the mechanisms of these processes are so far little understood and
 pose a formidable challenge. Attempts were made to study   the self assembly
 in different frameworks like `tile based self-assembly' \cite{BW,L,Win,Win1,Win2}.
 Perhaps the best model for self assembly was proposed by \cite{Win2}.
 With an aim of making the process of self assembly more clear, studies
 of abstract models,  such as self assembly of strings was initiated
 \cite{CPV}. In \cite{bot} authors introduce an operation among strings
 and languages, called ``superposition'', which is similar to the
 Csuhaj-Varj\'{u}'s operation called self assembly on strings, but
 their approaches are different.

\par
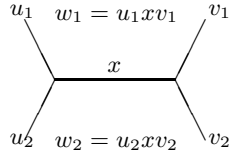
\begin{figure}[h]
\begin{center}
\setlength{\unitlength}{2mm}
\begin{picture}(14,10)(-7,-5)
\put(-4,0){\line(1,0){8}}
\put(-6,4){\line(1,-2){2}}
\put(-6,-4){\line(1,2){2}}
\put(6,4){\line(-1,-2){2}}
\put(6,-4){\line(-1,2){2}}
\put(-7,-4.3){$u_2$}
\put(-7,4.3){$u_1$}
\put(6.2,-4.3){$v_2$}
\put(6.2,4.3){$v_1$}
\put(-0.5,0.5){$x$}
\put(-4,4){$w_1=u_1xv_1$}
\put(-4,-4.5){$w_2=u_2xv_2$}
\end{picture}
\end{center}
\caption{A scheme for \sa of two strings}
\label{eqi}
\end{figure}

 Inspired by the different models of self-assembly, in particular the string
self assembly of Csuhaj-Varj\'{u} \cite{CPV}, we planned to propose
a string based operation which may be a generalisation of
self-assembly operation proposed in  Csuhaj-Varj\'{u}'s paper
\cite{CPV}.  In Csuhaj-Varj\'{u}'s model, two strings $uv$ and $vw$
self assemble over $v$ and generate $uvw$. Here $v$ is the
overlapping string. Then comes the question : What will be the
process if we do not restrict the overlapping
 string to be in the end of the first string and in the beginning of the second string.
 As an answer to the above question we propose a new operation on two strings.
 Two strings $u_1xv_1$
  and $u_2xv_2$ self assemble over the substring $x$ (also called
  overlapping string, $x\neq\ve$)
    and generate the strings $u_1xv_2$ and $u_2xv_1$ as illustrated
    in figure \ref{eqi}.

Normally, in any self-assembly process, no portion of the components
(that take part in the self-assembly) should be lost.  In that
sense, our new operation on strings (where some portions of the
strings are lost) can no longer be called as the abstraction of the
self-assembly process.
\par
But, our operation resembles in one sense, the recombination process
of chromosomes by exchanging the segments between homologous
chromosomes, called crossing-over. A chromosome is a single piece of
DNA that contains many genes, regulatory elements and other
nucleotide sequences.  Each gene occupies a well-defined site or
locus in its chromosome, having corresponding locations in the pair
of homologous chromosomes.  Chromosomal cross over (or crossing
over)is the process by which two chromosomes pair up and exchange
their DNA.Crossover usually occurs when matching regions on matching
chromosomes (homologous chromosomes) break and then reconnect to the
other chromosomes.  The result of this process is an exchange of
genes, called genetic recombination, which leads to the genetic
variability.  Crossover can occur at one or more points along the
adjacent chromosomes.
\par
In \cite{mitrana1}, an operation on strings and languages having the
same feature is introduced.  Every chromosome is considered as a
string.  The operation is applicable to a pair of strings of equal
length as the crossing over is between the homologous chromosomes.
\par
Each string is cut in several fragments, but in the sites for both
of them and crossing these fragments by ligases.  A new string, of
the same length, is formed by starting at the left end of one
parent, copying a segment, crossing over to the next site in the
other parent, copying a substring, crossing back to the first parent
and so on until the right end of one parent is reached.  Obviously,
another new string can be obtained by starting with the other
parent.  This crossover operation \cite{mitrana1} among the strings
is similar to the chromosome crossing-over.  A generalisation of the
splicing system is proposed in \cite{mitrana2}.
\par
Our proposal, two strings $u_1xv_1$
  and $u_2xv_2$ overlap at  the substring $x$
    and generate the strings $u_1xv_2$ and $u_2xv_1$, differs with
    the cross-over operation in two aspects. First, in our model,
    words of different lengths can participate in a crossover.
    Second, crossing over occurs at only one site between the words.
     For these reasons, we call our operation as
     {\em Generalised Sequential Cross Over (GCSO)}.  We use the
     adjective {\it generalised} in the sense that crossover can
     occur between any two words of any length and the adjective
     {\it sequential} in the sense that the crossover occurs between
     any two words at only one point(site) in contrary to the
     occurrence at one or more points between the chromosomes.

\par Any two strings may share more than one common overlap and so the result of GSCO
 of two strings is in general a set of strings. As usual in formal language theory,
  we extend GSCO to a language, iterated version of GSCO over a language.

\par Our study answers several questions in the sense of nano-scale fabrication;
like - can we decide if a given language can be obtained by iterated
GSCO and if so
 can we effectively construct a minimal finite set of initial strings. Given such
 a finite set of strings, what language can be generated by the
 GSCO?

\par Though the operation GSCO is just an abstraction of the  \sa operation introduced
 in \cite{mitrana1}, our study reveals many interesting results such as: iterated GSCO
  of any
 language will always be regular, and a subclass of GSCO languages matches exactly with
 the strictly locally testable
 language(SLT) \cite{Mc} leading to a new characterisation of SLT language using
 iterated GSCO.

\par Section 2 deals with the preliminaries required for this paper.
Section 3 introduces the GSCO operation on words and languages along
with some basic results. Section 4 discusses a variant of GSCO.
Section 4 shows that the operations 1-GSCO and 2-GSCO over a
language $L$ are the same. Two types of iterations are defined for
GSCO and their equivalence is discussed in section 5. Section 6
discusses the regularity of GSCO languages. Section 7 compares the
GSCO languages with the other regular subclasses.
\section{Preliminaries}
\par Throughout this paper, we assume that the reader is familiar with
the fundamental concepts of formal language theory and automata, i.e.
 notations of grammar and finite automata \cite{HMU}.
We list here some notations and notions we use in this paper.
\subsection{Basic notations  of formal language theory}
\par An alphabet is always a finite set of letters denoted by $\sg$. The set of
all words over an alphabet $\sg$ is denoted by $\sg^*$. The empty
word is denoted by $\ve$. Further $\sg^+=\sg^*\backslash\ve$. Given
a word $w$, the number of
 symbols in $w$ is the length of the word and is denoted by $|w|$. A word $v$
  is a sub-word (in literature, it is also called as {\it factor}) of  a word
   $w$ if there are words $u_1$ and $u_2$ (possibly empty) such that $w=u_1vu_2$.
    $v$ is called prefix  of $w$ ($\p(w)$) if $w=vu$. Similarly $v$ is called the
     suffix of $w$ ($\s(w)$) if $w=uv$. $\p(L)=\{\p(w):w\in L\}$  and
     $\s(L)=\{\s(w):w\in L\}$. The notation $\sg_x$ means the set of
     symbols of $\sg$ that
occurs in the word $x$.  $u_x$ means the word $u$ which is a sub-string
 of a word $x$. $|u|_x$ is the number of occurrence of $x$ in $u$.
  For a fixed $x$ (which is a sub-string of $u$), $|u|_x^r$
  is the total number of the occurrence of $x$ to the right of $x$.
  We define a function $I_x$ over the $\sub(x)$ such that,

\begin{eqnarray*}
I_x&:&\sub(x)\longrightarrow N\\
 I_x(u)&=&|u|_x-|u|_x^r.
\end{eqnarray*}
The class of regular language is defined by REG. Every finite automaton
induces a right invariant equivalence relation defined on the set of input
 strings which is formalised in the following theorem (see \cite{HMU})

\begin{theorem}[Myhill-Nerode]
The following statements are equivalent.
\begin{enumerate}
\item The set $L\subseteq\sg^*$ is accepted by some finite automaton.
\item $L$ is the union of some of the equivalence classes of a right
invariant equivalence relation of finite index.
\item Let equivalence relation $R_L$ be defined by $xR_Ly$ if and only if
 $\forall z\in\sg^*,~xz\in L$ exactly when $yz\in L$. Then $R_L$ is of finite index.
\end{enumerate}
\end{theorem}
\subsection{Splicing}
\par A splicing rule (over alphabet $\sg$) is a quadruple $(u_1,u_2,u_3,u_4)$
of words $u_1,u_2,u_3,u_4\in\sg^*$  which is often written as
follows: $u_1\#u_2\$u_3\#u_4$. Here $\#$ and $\$$ are splicing
symbols which are not in $\sg$. A splicing rule
$r=u_1\#u_2\$u_3\#u_4$ is applicable to two words $x=x_1u_1u_2x_2$
and $y=y_1u_3u_4y_2$. The splicing of the words $x$ and $y$ by the
splicing rule $u_1\#u_2\$u_3\#u_4$, produces two new words
$w_1=x_1u_1u_4y_2$ and $w_2=y_1u_3u_2x_2$. In this case we write
$(x,y)\vdash_r(w_1,w_2)$. This operation is also called 2-splicing.
We can take only $w_1$ as a result instead of both of them. In that
case the corresponding operation is called 1-splicing and is denoted
by $(x,y)\vdash w_1$.

\par A pair $\sigma=(\sg,R)$ where $\sg$ is an alphabet and $R$ is a set of
splicing rules is called a splicing scheme or a $H$-scheme. For an $H$-scheme
$\sigma=(\sg,R)$ and a language $L\subseteq\sg^*$, we define
\[\sigma(L)=\{w_1,w_2\in\sg^*|x,y\in L,~r\in R,~ (x,y)\vdash_r w_1,w_2\}\]
where $x,y,w_1,w_2$ and $r$ are specified above. The iterative version of the
splicing operation is defined as
\begin{eqnarray*}
&\sigma^0(L)=L\\
&\sigma^{i+1}(L)=\sigma^i(L)\cup\sigma(\sigma^i(L))\\
&\sigma^*(L)=\bigcup_{i\geq0}\sigma^i(L)
\end{eqnarray*}

\par $H$-system is a construct $H=(\sg,A,R)$ where $\sg$ is a finite alphabet,
$A\subseteq \sg^*$  is a set of initial words over $\sg$, called
axiom and $R\subseteq \sg^*\#\sg^*\$\sg^*\#\sg^*$ is a set of
splicing rules. The language generated by   $H=(\sg,A,R)$ is
$\sigma^*(A)$. Thus the language generated by the $H$-system is the
set of all words that can be generated starting with $A$, as initial
words and by iteratively applying splicing rules from $R$ to the
words already generated.

\par A $H$-system is called a `null context \h-system' (NCH) if   $R$ is a
finite subset of $\sg^*$. The language generated by NCH is the
smallest language $L$ in $\sg^*$ that contains $A$ and has the
property that whenever strings $wrx$ and $yrz$ are in $L$, $r\in R$;
the strings $wrz$ and $yrx$ are also in $L$. A language $L$ is
called a null context splicing language (NC\h-language) if there
exists a null context splicing system that generates $L$
\cite{head}. Simple \h-system  \cite{MPS} is a \h-system
$(\sg,A,R)$, where $R\subseteq\sg$ such that for $x,y,z\in\sg^*$and
$a\in R;~(x,y)\vdash^az$ if and only if
$x=x_1ax_2,~y=y_1ay_2,~z=x_1ay_2$, for $x_1,x_2,y_1,y_2,a\in\sg^*$.
The family of simple \h-systems is a subclass of NCH systems. $SH$
is the family of languages generated by a simple splicing system.

\subsection{Constant}
\par The concept of a constant, as introduced by Schutzenberger \cite{swat}
 is a valuable conceptual tool for splicing theory, given out many years before
  the proposal  of the theory of splicing. A string $c\in \sg^*$ is a constant for  a
  language $L$ over an alphabet $\sg$ if, whenever $wcx$ and $ycz$ are in $L$,
   both $wcz$ and $ycx$ are also in $L$. A string $y$ is a factor of a string
   $w$ if $w=xyz$ for some $x,y\in\sg^*$ and that $y$ is a factor of a language
    $L$ if $y$ is a factor of some string in $L$. Further each rule of a NCH
     system $G$ is necessarily a constant for the language $L(G)$.

  \subsection{Strictly locally testable languages}
  The concept of strictly locally testable languages   was introduced
  by McNaughton and Papert in \cite{Mc}.  Later, De Luca and Restivo
  \cite{Luc} gave a characterisation for such languages, using  the concept of
  constants \cite{swat}. We give the definition of strictly locally testable languages as in
  \cite{Mc} and the characterisation of it as in \cite{Luc}.
  \begin{defin}\cite{Mc}
  A subset $X $of $A^+$ is called strictly locally testable if a
positive integer $k$ and three subsets $U, V, W $of $A^k$  exist
such that:$ X \cap A^kA^* = (UA^* \cap A^*V)\setminus A^*WA^*$.
  \end{defin}
  Class of strictly locally testable languages is denoted by $SLT$
\begin{defin}{Characterisation of SLT \cite{Luc}:}
 A Language $L$ is a SLT if there is a positive integer $k$ for
which every factor of $L$ of length $k$ is a constant.
\end{defin}

\section{Generalised Sequential Crossover}
\begin{defin}
Generalised sequential \sa scheme $GSCO=(\sg,R)$, where $\sg$ is the
finite alphabet,
 $R\subseteq \sg^*$ be the finite set of overlapping strings; we write $GSCO=(\sg,R)$
 as $GSCO_R$. $GSCO_R$ is also called a $R$-crossover. When $R$ is singleton,
  say $R=\{x\}$, we write $GSCO_x$ instead of $GSCO_R$.

\par For a given GSCO scheme $GSCO$ and two words $w_1=u_1xv_1$ and $w_2=u_2xv_2\in \sg^*$,
 we define
\[GSCO_x(w_1,w_2)=\{u_1xv_2,u_2xv_1\in\sg^*:  w_1=u_1xv_1, w_2=u_2xv_2,~
\ve\neq x\in R\}.\]
The scheme is shown in figure \ref{eqi}.
\end{defin}
\par Instead of writing $GSCO_x(u_1xv_1, u_2xv_2)$, we also write
 $u_1xv_1>^x\rs u_2xv_2=\{ u_1xv_2,u_2xv_1\}$, which means that the two strings
  $u_1xv_1$ and $u_2xv_2$ \sae over the sub-string $x$ to generate two new
  words  $u_1xv_2$ and $u_2xv_1$. We also write $u_1xv_1>^x\rs u_2xv_2=\{u_1xv_2,u_2xv_1\}$
   instead of $(u_1xv_1,u_2xv_2)>^x\rs\{u_1xv_2,u_2xv_1\}$. Then
\[GSCO_R(w_1,w_2)=\bigcup_{x\in R}w_1>^x\rs w_2.\]
  Obviously $R$ should contain words which are sub-words found in both
  $w_1$ and $w_2$, otherwise $GSCO_R(w_1,w_2)$ will be empty. We call the operation
  $GSCO_x,~x\in\sg$ as the symbol overlapping GSCO. Similarly we call $GSCO_x,~x\in\sg^*$
  as the string overlapping GSCO. Let $\sub(w)$ be the set of all sub-words of $w$.
   If in a GSCO scheme $R=\sub(w_1)\cap\sub(w_2)$, we simply write $GSCO(w_1,w_2)$, i.e.
    $GSCO(w_1,w_2)$ is the set of all words that can be generated by the GSCO
     of $w_1$ and $w_2$ with all possible overlapping. In other words,
\[GSCO(w_1,w_2)= \bigcup_x GSCO_x(w_1,w_2),~~~~x\in\sub(w_1)\cap\sub(w_2).\]
We do not \sae two strings with $\ve$ as the overlapping string.\footnote{Out
of curiosity we record the result
\[GSCO_\ve(w_1,w_2)=\mathrm{Pref}(w_1).\mathrm{Suff}(w_2)\cup \mathrm{Pref}(w_2).
\mathrm{Suff}(w_1).\]}
\par We extend the above definition  to languages. Given any two languages $L_1$ and $L_2$ over the alphabet $\sg_1$ and $\sg_2$ respectively such that $\sg_1\cap\sg_2\neq\emptyset$, we define
\[GSCO_R(L_1,L_2)=\bigcup_{\substack{w_1\in L_1\\w_2\in L_2}}GSCO_R(w_1,w_2).\]
Here the underlying crossover scheme is $GSCO=(\sg_1\cup\sg_2,R)$.
As mentioned earlier, when  $R=\sub(L_1)\cap\sub(L_2)$ ($R$ is the
set of all possible overlapping between a word of \(L_1\) and a word
of \(L_2\).
\[GSCO_R(L_1,L_2)=\bigcup_{\substack{w_1\in L_1\\w_2\in L_2}}GSCO(w_1,w_2).\]
\(GSCO(L,L)\) is written as just \(GSCO(L)\).

\par We record some results, whose proofs are immediate.
\begin{prop}
Let $u,v\in\sg^*$.
\begin{enumerate}
\item $GSCO_x(u,v)=GSCO_y(u,v)$ where the sub-word $x$ occurs in $y$ only once
and no two  symbols of  $x$ are same.
\item $GSCO_x(u,v)\supseteq GSCO_y(u,v)$, $x$ is a sub-word of $y$.
\item If $R\subseteq R',~GSCO_R(u,v)\subseteq GSCO_{R'}(u,v)$.
\item $GSCO_{R_1\cup R_2}(u,v)=GSCO_{R_1}(u,v)\cup GSCO_{R_2}(u,v)$.
\item $GSCO_{R_1\cap R_2}(u,v)=GSCO_{R_1}(u,v)\cap GSCO_{R_2}(u,v)$.
\item $GSCO_a(GSCO_a(u,v),u)=GSCO(u,GSCO_a(u,v)),~a\in\sg$.
\item $GSCO_a(GSCO_a(u,v),v)=GSCO(v,GSCO_a(u,v)),~a\in\sg$.
\item If $GSCO_a(u,v)=\{x,y\}$, then $GSCO_a(x,y)=\{u,v\}$, i.e.
the operation $GSCO_a,~\forall a \in\sg$ is reversible.
\item The length of the words in $GSCO(u,v)$ will range form 1 to $|u|+|v|-1$.
\item $GSCO(w,w)=w$ if no two  symbols of $w$ are same.
\item GSCO operation is not associative over words, but commutative over words.
In fact
$GSCO(L_1,L_2)=GSCO(L_2,L_1)$.
\item $GSCO(a^i,a^j)=\{a,a^2,\cdots,a^{i+j-1}\}$.
\item For any two languages, $L_1$ and $L_2$
\[GSCO(L_1\cup L_2)=GSCO(L_1)\cup GSCO(L_2)\cup GSCO(L_1,L_2).\]
\item $GSCO(w,w^R)=\{uau^R:u\in\p(w), ~a\in\sg\}$.
\item \label{1}For any two words $w_1,w_2\in\sg^*$, and
$x\in\sub(w_1)\cap\sub(w_2)$,
\[GSCO_x(w_1,w_2)\subseteq GSCO_{a\in\sg_x}(w_1,w_2).\]
If a word is generated by a string overlapping ($x$ overlapping)
  GSCO of $w_1$ and $w_2$, then the word can also be
generated by a symbol (that occurs in $x$) overlapping.
\end{enumerate}
\end{prop}

\begin{proof}
All but the last of the above statements follows directly from the
definition. We only prove the last one (statement \ref{1}).
Let $u\in GSCO_x(w_1,w_2),~x\in\sub(w_1)\cap\sub(w_2)$. If $x\in\sg$,
then the proof is immediate.\\
Let $x\notin\sg,~|x|\geq2$. Let $x=a_1a_2\cdots a_n$, where some
$a_i$'s
 may be same. Suppose $w_1=u_1 a_1a_2\cdots a_nv_1, ~w_2=u_2 a_1a_2\cdots a_nv_2$.
 Then $u\in\{u_1 a_1a_2\cdots a_nv_2,u_2 a_1a_2\cdots a_nv_1\}$.
\begin{description}
\item[Case I] $u=u_1 a_1a_2\cdots a_nv_2.$
\[u\in GSCO_{a_i}(w_1,w_2),i\in\{1,2,\cdots,n\}~\Rightarrow ~u\in
GSCO_{a\in\sg_x}(w_1,w_2).\]
 Hence $GSCO_x(w_1,w_2)\subseteq
GSCO_{\sg_x}(w_1,w_2)$.
\item[Case II]  $u=u_2a_1a_2\cdots a_nv_1$. We get the result similarly. Hence the proof.
\end{description}
\end{proof}
\begin{note}
The other way of the statement \ref{1} is not true, i.e.
\[ GSCO_{a\in\sg_x}(w_1,w_2)\not\subseteq GSCO_{x}(w_1,w_2).\]
As an example: $GSCO_{aba}(c_1abac_2,d_1abad_2)=\{c_1abad_2,d_1abac_2\}$.
 But $GSCO_a(c_1abac_2,d_1abad_2)=$ \\
$\{c_1abad_2,d_1abac_2,c_1ad_2,d_1ababac_2,c_1ababad_2,d_1ac_2\}$.
\end{note}

\begin{example}
$GSCO(\{a^n:n\geq1\})=a^+$.
\end{example}

\begin{example}
$GSCO(\{a^nb^n:n\geq1\})=a^+b^+$.
\end{example}

\begin{example}
Let $L=\{ab,ba,bb\}$.
\begin{align*}
ab \ers ab &=\{ab\} &ab\ers ba &=\{a,b,bab,aba\}\\
ab\ers bb &=\{ab,bb,b,abb\} &ba\ers bb &=\{b,bb,ba,bba\}\\
ba\ers ba &=\{ba\} &bb\ers bb &=\{bb\}.
\end{align*}
So we have
\[GSCO(L)=\{a,b,ab,ba,bb,aba,bab,abb,bba\}.\]
\end{example}

\begin{example}
$GSCO(\{a,b\})=\{a,b\}$.
\end{example}

\begin{example}
$GSCO(\{abcab,c\})=\{ab,abc,cab,abcabcab\}$.
\end{example}

\par In computing $GSCO (w_1,w_2)$, one has to first compute all the common sub-strings $x$
and compute $\bigcup_xGSCO_x(w_1,$ $w_2)$. For $GSCO(L)$  we have to
compute $\bigcup_{w_1,w_2\in L}GSCO(w_1,w_2)$. In short,
\[GSCO(L)=  \bigcup_{w_1,w_2\in L}~\bigcup_xGSCO(w_1,w_2),\hspace{3mm}x\in
\sub(w_1)\cap\sub(w_2)
,\]
which increases the complexity of the computation of GSCO. We
have the following theorem to reduce this tedious calculation of
finding all the common sub-strings of all the pairs of  words of a
given language $L$.
\begin{theorem}
Let $w_1,w_2\in\sg^*$.
\[GSCO(w_1,w_2)=\bigcup_{a\in\sg_{w_1}\cap\sg_{w_2}}GSCO_a(w_1,w_2).\]
\end{theorem}
\begin{proof}
Since
\[GSCO(w_1,w_2)= \bigcup_{x\in\sub(w_1)\cap\sub(w_2)} GSCO_x(w_1,w_2),\]
it is enough if we prove that:
\[ \bigcup_{x\in\sub(w_1)\cap\sub(w_2)} GSCO_x(w_1,w_2)=\bigcup_{a\in\sg_{w_1}
\cap\sg_{w_2}}GSCO_a(w_1,w_2).\]
\par Since $\sg_{w_1}\cap\sg_{w_2}\subseteq\sub(w_1)\cap\sub(w_2)$,
\[GSCO_{\sg_{w_1}\cap\sg_{w_2}}(w_1,w_2)\subseteq
GSCO_{\sub(w_1)\cap\sub(w_2)}(w_1,w_2).\]

\[ \bigcup_{a\in\sg_{w_1}\cap\sg_{w_2}} GSCO_a(w_1,w_2)
\subseteq\bigcup_{x\in\sub(w_1)\cap\sub(w_2)}GSCO_x (w_1,w_2).\]

\par To prove the other way, let  $u\in GSCO_x(w_1,w_2)$. If
$x\in\sg_{w_1}\cap\sg_{w_2}$, then the proof is obvious. Suppose
$|x|\geq2$ (i.e. $x$ is a common sub-string  of $w_1$ and $w_2$). By
the result \ref{1} of proposition 1,  there exists a symbol in $x$,
say $a$, (i.e. $a\in\sg_x$) such that $u\in GSCO_a(w_1,w_2)$. Since
$a\in\sg_x, ~x\in\sub(w_1)\cap\sub(w_2)$, we have
$a\in\sg_{w_1}\cap\sg_{w_2}$. This implies $u\in
GSCO_{a\in\sg_{w_1}\cap\sg_{w_2}}(w_1,w_2)$. Hence
\[GSCO_x(w_1,w_2)\subseteq GSCO_{a\in\sg_{w_1}\cap\sg_{w_2}}(w_1,w_2).\]

\[\Rightarrow \bigcup_{x\in\sub(w_1)\cap\sub(w_2)}GSCO_x(w_1,w_2)
\subseteq \bigcup_{a\in\sg_{w_1}\cap \sg_{w_2}}GSCO_a(w_1,w_2).\]
\end{proof}
\begin{corollary}\label{c1}
$GSCO(w_1,w_2)=\bigcup_{a\in\sg}GSCO_a(w_1,w_2)$.
\end{corollary}
\begin{proof}
It is enough if we prove that
\[\bigcup_{a\in\sg_{w_1}\cap\sg_{w_2}}GSCO_a(w_1,w_2)=\bigcup_{a\in\sg}GSCO_a(w_1,w_2).\]
The alphabet
\begin{equation}\label{e1}
\sg=(\sg_{w_1}\cap\sg_{w_2})\cup A,
\end{equation}
where $A$ contains the symbols of $\sg$ which are not in
$\sg_{w_1}\cap\sg_{w_2}$, i.e. the alphabet $\sg$ can be written as
a disjoint union of the two sets with respect to the words $w_1$ and
$w_2$.
\begin{equation}
\bigcup_{a\in
A}GSCO(w_1,w_2)=\bigcup_{a\notin\sg_{w_1}\cap\sg_{w_2}}GSCO(w_1,w_2)=\emptyset.
\end{equation}
By result 4 of proposition 1, (\ref{e1}) implies
\begin{eqnarray*}
\bigcup_{a\in\sg}GSCO_a(w_1,w_2)&=&\big(\bigcup_{a\in\sg_{w_1}\cap\sg_{w_2}}
GSCO(w_1,w_2)\big)\bigcup\big(\bigcup_{a\in A}GSCO_{a\in A}(w_1,w_2)\big)\\
\Rightarrow\bigcup_{a\in\sg}GSCO_a(w_1,w_2)&=&\bigcup_{a\in\sg_{w_1}\cap
\sg_{w_2}}GSCO(w_1,w_2)\big).
\end{eqnarray*}
Hence the proof.
\end{proof}
\begin{corollary}
$GSCO(L)=\bigcup_{w_1,w_2\in L}\bigcup_{a\in\sg}GSCO_a(w_1,w_2).$
\end{corollary}
\begin{proof}
\begin{eqnarray*}
GSCO(L)&=&\bigcup_{w_1,w_2\in L}GSCO(w_1,w_2)\\
    &=&\bigcup_{w_1,w_2\in L}\bigcup_{a\in\sg}GSCO_a(w_1,w_2).
\end{eqnarray*}
\end{proof}
\par This corollary tells us that to compute $GSCO(L)$ it is enough to compute the GSCO of
 $w_1$ and $w_2$ over the symbols of the alphabet $\sg$ and take the union of all those
 $ GSCO(w_1,w_2)$'s.

\subsection{CGSCO}
\par We mention a special type of the operation GSCO  viz.,  Corresponding GSCO (CGSCO).

\begin{defin}[CGSCO]
Given any two words $w_1,w_2$, and let $x$ be a common sub-string of
them such that in both $w_1$ and $w_2,~x$ occurs more than once. We
\sae $w_1$ and $w_2$ in such a way that the first occurrence of $x$
in $w_1$ overlaps with the   first occurrence of $x$ in $w_2$
(second occurrence in $w_1$ \sae with second occurrence of $x$ in
$w_2$ and so on). We call such a GSCO as Corresponding GSCO.
\end{defin}
\par As an example $CGSCO(\underline{ab}c\underline{ab},
\underline{ab}\hspace{0.5mm}\underline{ab})=\{ab,abab,abcab\}$. The
sub-strings which occurs in both the strings more than once are
$ab,a , b$.
 Here we do not allow the overlap of the first occurrence of $ab$ in  $abcab$ with
 second occurrence of $ab$ in $abab$.
\par As seen in proposition 1, result \ref{1} we have
\[GSCO_x(w_1,w_2)\subseteq GSCO_{a\in\sg_x}(w_1,w_2).\]
There are some GSCO's for which the equality holds; i.e.
 for every symbol overlapping GSCO  of $w_1$ and $w_2$,
 there exists a string overlapping GSCO of  $w_1$ and $w_2$.
  If $x$ is a common sub-string in $w_1$ and $w_2$, then any
  sub-string of $x$ is also a common string, GSCO  can occur
  by the overlapping of the sub-string of $x$ also. Result 1
  of the proposition 1 tells that   $GSCO_x(w_1,w_2)\supseteq GSCO_y(w_1,w_2)$
  where $x\subseteq y$. To compute the $GSCO(w_1,w_2)$ we have to consider
   all the possible common sub-strings. But for the GSCO  systems, which
    satisfies the property $GSCO_{a\in\sg_x}(w_1,w_2)=GSCO_x(w_1,w_2)$, of
    theorem 2. To calculate $GSCO(w_1,w_2)$ it is enough to compute
    $GSCO_{a\in\sg_x}(w_1,w_2)$ where $x$ is the maximal common sub-string
    of $w_1$ and $w_2$ (A common sub-string $x$ is said to be maximal if there is no
     common sub-string $y$ such that $x$ is  a sub-string of $y$), i.e.
\[GSCO(w_1,w_2)=\bigcup_x GSCO_{a\in\sg_x}(w_1,w_2),\]
 where $x$ is the common maximal sub-string of $w_1$ and $w_2$.
\begin{theorem}
A GSCO is a CGSCO if and only if
\[GSCO_x(w_1,w_2)=\bigcup GSCO_{a|x}(w_1,w_2)\]
Here $a_x$ is any symbol from the sub-string $x$ such that
$I_{w_1}(x)=I_{w_2}(x)$. $GSCO_{a|x}$ is an operation where the
overlapping occurs over $a$ which is a sub-string of $x$ ans not
elsewhere.
\end{theorem}
\begin{proof}
Let the GSCO be a CGSCO. Let $w_1$ and $w_2$ be any two words. Let
$x$ be a common sub-string of  $w_1$ and $w_2$. $|w_1|_x=|w_2|_x=n$.
Since the GSCO is a CGSCO, $w_1$ and $w_2$ can \sae over $x$ only
for $n$ times.  Let $x$ occurs $n$ times in $w_1$ and $m$ times in
$w_2$.
\par In the calculation of $GSCO_x(w_1,w_2)$ we have to consider all the possible overlapping of $x$, i.e. any $x$ in $w_1$ can overlap with any $x$ in $w_2$. Let $x=a_1a_2\cdots a_k,~w_1=u_1xu_2xu_3\cdots xu_{n+1},~w_2=v_1xv_2xv_3\cdots xv_{m+1}$. We have assumed that GSCO is a CGSCO. Moreover, we have to consider such $x$ overlapping such that $I(x_{w_1})=I(x_{w_2})$, i.e. we calculate $GSCO_x(w_1,w_2)$ when the $i$th occurrence of $x$ in $w_1$ overlaps with the $i$th occurrence of $x$ in $w_2$.  In such a case,
\begin{equation}\label{m1}
GSCO_x(w_1,w_2)=\{u_1xv_2xv_3\cdots xv_{n+1},v_1xu_2xu_3\cdots
xu_{n+1};
        \cdots;u_1xu_2xv_3\cdots u_n xv_{n+1},v_1xv_2xv_3\cdots v_nxu_{n+1}\}.
\end{equation}
We consider the sub-string $x$ in $w_1$ and sub-string $x$ in $w_2$ such that $I(x_{w_1})=I(x_{w_2})$. This means if we consider $x$ which occurs $i$th time in $w_1$ we have to \sae it with the $i$th occurrence of $x$ in $w_2$ as a sub-string.
\par Consider $x$ such that $I_{w_1}(x)=I_{w_2}(x)=1$, i.e. the $x$ which occurs first time in $w_1$ as well as in $w_2$.Let $a$ be any symbol in the sub-string $x$, $a=a_l$ say.

\par By hypothesis GSCO is  a CGSCO. We compute $CGSCO_{a|x}(w_1,w_2)$. $CGSCO_{a|x_i}(w_1,w_2)$ means that the overlapping occurs between the $i$th occurrence of $a$ in $x$ which occurs in $w_1$ and the $i$th occurrence of $a$ in $x$ which occurs in $w_2$. In our case, if $a$ is the $i$th symbol in $x_{w_1}$ then $a$ is also the $i$th symbol in $x_{w_2}$. $a$ can occur many times in $x$, but the overlapping of $a$ has to take place in the corresponding position for $CGSCO_{a|x}$.
\begin{eqnarray*}
 CGSCO_{a|x}(w_1,w_2)&=&CGSCO_{a|x}(u_1a_1\cdots a_ku_2x\cdots xu_{n+1},v_1a_1\cdots a_kv_2x\cdots xv_{n+1})\\
        &=&\{u_1a_1\cdots a_la_{l+1}a_kv_2x\cdots xv_{n+1},v_1a_1\cdots a_la_{l+1}a_ku_2x\cdots xu_{n+1}\}\\
        &=&\{u_1xv_2x\cdots xv_{n+1},v_1xu_2x\cdots xu_{n+1}.\}
\end{eqnarray*}
 If $a=a_j$, and  the \sa occurs over $a_j$ in $x_{w_1}$ and   $a_j$ in $x_{w_2}$. The calculation is similar, and we get
\begin{eqnarray*}
 CGSCO_{a|x}(w_1,w_2)&=&CGSCO_{a|x}(u_1a_1\cdots a_ku_2x\cdots xu_{n+1},v_1a_1\cdots a_kv_2x\cdots xv_{n+1})\\
        &=&\{u_1a_1\cdots a_ja_{j+1}a_kv_2x\cdots xv_{n+1},v_1a_1\cdots a_ja_{j+1}a_ku_2x\cdots xu_{n+1}\}\\
        &=&\{u_1xv_2x\cdots xv_{n+1},v_1xu_2x\cdots xu_{n+1}.\}
\end{eqnarray*}
It does not matter, how many times $a$ is repeated in $x$, as the \sa is taking place on its position of occurrence (in the sub-string $x$ of both the words) only.

\par We repeat the case I for $x$ such that $I(x_{w_1})=I(x_{w_2})=2$. Arguing on similar line,
\[GSCO_{a|x}(w_1,w_2)=\{u_1xu_2xv_3\cdots xv_{n+1},v_1xv_2xu_3\cdots xu_{n+1}\}.\]
Similarly we have
\[GSCO_{a|x}(w_1,w_2)=\{u_1xu_2x\cdots u_ixv_{i+1}\cdots xv_{n+1},v_1xv_2x\cdots v_i\cdots xu_{i+1}\cdots xu_{n+1}\}~\mbox{where~}I(x_{w_1})=I(x_{w_2})=2.\]
So we have
\begin{equation}\label{m2}
\begin{split}
\bigcup_{\substack{I{w_1}(x)=I_{w_2}(x)\\=1}}^jGSCO_{a|x}(w_1,w_2) & =\{u_1xu_2x\cdots u_jxv_{j+1}\cdots xv_{n+1},\\
    & \quad v_1xv_2x\cdots v_j\cdots xu_{j+1}\cdots xu_{n+1}: j=1,2,\cdots \min\{|w_1|_x,|w_2|_x\}\}.
\end{split}
\end{equation}
By (\ref{m1}) and (\ref{m2}), we have the claim.\\
\par Given
\[GSCO_x(w_1,w_2)=\bigcup_{\substack{I(x_{w_1})=I(x_{w_2})\\=2}}^jGSCO_{a:x}(w_1,w_2),\]
to show that GSCO is a CGSCO.\\
\par Let the above claim be not true, i.e. GSCO is not a CGSCO. Choose $w_1=u_1xu_2xu_3\cdots xu_{n+1}$ and $w_2=v_1xv_2xv_3\cdots xv_{n+1}$. As we have noted earlier, the number of $x$-overlapping for a CGSCO  depends on the minimum number of occurrences of $x$ in the two words to be self-assembled. Hence, without loss of generality we may assume that both of them has the same number of $x$ as a sub-word.
\par When the first occurrence of $x$ in $w_1$ overlaps with the third occurrence of $x$ in $w_2$, we get two new words
\begin{equation}\label{m3}
u_1xv_4xv_5\cdots v_{n+1},v_1xv_2xv_3xu_2xu_3\cdots u_{n+1}\in
GSCO_x(w_1,w_2).
\end{equation}
The above strings can only be generated by $GSCO_{a|x}(w_1,w_2)$
where $I_{w_1}(x)=1$ and $I_{w_2}(x)=3$. It can not be generated by
$GSCO_{a|x}(w_1,w_2)$  where $I{w_1}(x)=I_{w_2}(x)$. Hence
\begin{equation}\label{m4}
u_1xv_4xv_5\cdots v_{n+1},v_1xv_2xv_3xu_2xu_3\cdots u_{n+1}\notin
\bigcup_{\substack{I{w_1}(x)=I_{w_2}(x)\\=1}}^jGSCO_{a|x}(w_1,w_2).
\end{equation}
(\ref{m3}) and (\ref{m4}) contradicts our hypothesis. Hence the GSCO
is a CGSCO.
\end{proof}
\begin{corollary}
$CGSCO_x(w_1,w_2)=CGSCO_{y|x}(w_1,w_2)$, where $y$ is  a sub-string
of $x$.
\end{corollary}
\begin{proof}
The argument follows in the same line as in the previous theorem.
Since we are dealing with a CGSCO; the first $x$ of $w_1$ will match
with the first $x$ of $w_2$. Again in this also
\end{proof}

\section{1-GSCO and 2-GSCO}
  In the theory of splicing, two types of  splicing operations have been considered:
  the 1-splicing operation, when by applying a rule on two words, only one word is
  generated/considered; and the 2-splicing when both the two words are generated/considered.
\par In a similar line we introduce two operations: 1-GSCO and 2-GSCO.
The operations GSCO over the words $w_1$ and $w_2$ generate  two new
words,
 each  time  when $w_1$ and $w_2$   overlap over  a common sub-string $x$.
 For a common sub-string $x$, different overlaps are also possible.
 Collection of all such words is denoted by $GSCO_x(w_1,w_2)$. $GSCO(w_1,w_2)$
 is a collection of all possible $GSCO_x(w_1,w_2)$'s. $GSCO(L)$ is the collection of
  all $GSCO(w_1,w_2)$'s for all possible pairs of $w_1,w_2\in L$. Hence,
  the operation GSCO is made up of many `overlapping', with each overlapping
  generating two words.
\par The operation GSCO is called 1-GSCO if in all the concerned overlapping,
 we consider the word which has the prefix of the first word and the suffix of the
  second word as the only word generated. So $1GSCO_x(u_1xv_1,u_2xv_2)=\{u_1xv_2\}$,
  i.e. the operation 1GSCO generates only one word. We denote 1GSCO by $>_1\rs$.

\begin{figure}[h]
\begin{center}
\setlength{\unitlength}{1.5mm}
\linethickness{0.25cm}
\begin{picture}(14,10)(-7,-5)
\thicklines

\put(-4,0){\line(1,0){8}}
\put(-6,4){\line(1,-2){2}}
\put(6,-4){\line(-1,2){2}}
{\gray\put(-6,-4){\line(1,2){2}}
\put(6,4){\line(-1,-2){2}}}

\put(-7,-4.3){$u_2$} \put(-7,4.3){$u_1$} \put(6.2,-4.3){$v_2$}
\put(6.2,4.3){$v_1$} \put(-0.5,0.5){$x$}
{\red\put(6,-1){$1GSCO_x(u_1xv_1, u_2xv_2)=\{u_1xv_2\}$}}
\end{picture}
\end{center}
\caption{A scheme for 1GSCO of two strings. Scheme for the  output
string \(u_1xv_2\) is prominently shown. The grey part is the
discarded self-assembled string.} \label{eqi1}
\end{figure}
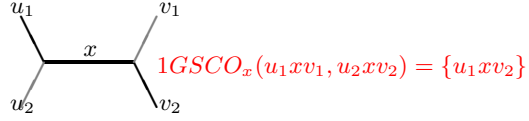
\par The operation GSCO is called 2GSCO if in all the concerned overlapping we consider
 both the words generated. So the operation 2GSCO coincides with GSCO.

\begin{lem}\label{mew}
For the two words $w_1,~w_2$
\begin{enumerate}
\item $1GSCO_x(w_1,w_2)=1GSCO_x(w_2,w_1)$, if and only of $w_1=w_2$.
\item $1GSCO(w_1,w_2)\subseteq2GSCO(w_1,w_2)$.
\item $1GSCO(w_1,w_2)\cup 1GSCO(w_2,w_1)=2GSCO(w_1,w_2)$.
\end{enumerate}
For any two languages $L_1$ and $L_2$;
\begin{enumerate}
\item[4.] $1GSCO(L_1,L_2)\subseteq2GSCO(L_1,L_2)$.
\item[5.] $1GSCO(L)=2GSCO(L)=GSCO(L)$.
\end{enumerate}

\end{lem}
\begin{proof}
The results 1, 2, 3 and 4 are obvious. We prove the result 5. When
the language $L$ is a singleton set, $1-GSCO(L)= 2-GSCO(L)$.
\begin{eqnarray*}
1GSCO(L)&=&\bigcup_{w_1,w_2\in L}1GSCO(\{w_1,w_2\})\\
    &=&\bigcup_{w_1,w_2\in L}\big(1GSCO(w_1,w_2)\cup1GSCO(w_2,w_2)\cup1GSCO(w_1,w_1)\cup1GSCO(w_2,w_2)\big)\\
    &=&\bigcup_{w_1,w_2\in L}\big(2GSCO(w_1,w_2)\cup2GSCO(w_1,w_1)\cup2GSCO(w_2,w_2) \big)\\
    &=&\bigcup_{w_1,w_2\in L}2GSCO(\{w_1,w_2\})\\
    &=&2GSCO(L).
\end{eqnarray*}
Since $2GSCO(L)$ is just $GSCO(L)$ we have the result.
\end{proof}
\par In case of finite H-system 1-splicing operation is more powerful than 2-splicing.
In GSCO system they coincide. By the  result 5 of Lemma \ref{mew}, to calculate $GSCO(L)$
it is enough to calculate $1GSCO(L)$, which is equivalent to  $GSCO(L)$. From now onwards
$GSCO(L)$ means either $1GSCO(L)$ or $2GSCO(L)$.
\section{Iterated GSCO}
\begin{defin}
Given a language $L$, we define the language obtained from $L$ by
unrestricted iterated application of GSCO. This language, called the
unrestricted GSCO closure of $L$, denoted by $uGSCO^*(L)$, is
defined as
\begin{eqnarray*}
uGSCO^0(L)&=&L\\
uGSCO^{i+1}(L)&=&uGSCO^{i}(L)\cup uGSCO(uGSCO^i(L))\\
uGSCO^*(L)&=&\bigcup_{i\geq0}uGSCO^{i}(L)
\end{eqnarray*}
\end{defin}

\par Clearly $uGSCO^*(L)$ is the smallest language containing $L$ and is closed under GSCO. That is, it is the smallest language $K$ such that $L\subseteq K$ and $GSCO(K)\subseteq K$. In other words, one starts with any pair of words in $L$ and apply GSCO iteratively to any pair of words previously produced. All the obtained words are collected.

\begin{defin}
For a word $w$ and a sub-string $x$ of $w$ we define the $\p_x(w)$, $\s_x(w)$ as follows:
\begin{eqnarray*}
\p_x(w)&=&\{u:uxu'=w;~u,u'\in\sg^*\}\\
\s_x(w)&=&\{s:s'xs=w;~s,s'\in\sg^*\}\\
\p_x(L)&=&\bigcup_{w\in L}\p_x(w)\\
\s_x(L)&=&\bigcup_{w\in L}\s_x(w).
\end{eqnarray*}
\end{defin}
It is clear that
\[w_1>_1^x\rs w_2=\p_x(w_1)\cdot x\cdot\s_x(w_2).\]
\begin{lem}\label{its1}
For any word $w$,
\[\p_x(\s_x(w))=\s_x(\p_x(w)).\]
\end{lem}
\begin{proof}
Let
\begin{eqnarray*}
&&u\in\s_x(\p_x(w)).\\
\Leftrightarrow&&\exists u'\in \sg^*~\mbox{such that~}u'xu\in\p_x(w)\\
\Leftrightarrow&&\exists u''\in \sg^*~\mbox{such that~}u'xuxu''=w\\
\Leftrightarrow&&uxu''\in\s_x(w)\\
\Leftrightarrow&&u\in\p_x(\s_x(w)).
\end{eqnarray*}
Hence the proof.
\end{proof}
\begin{lem}\label{its2}
For any three words $w_1,~ w_2,~w_3$
\[ (w_1>_1^x\rs w_2)>_1^x\rs w_3=w_1>_1^x\rs (w_2>_1^x\rs w_3);\]
i.e. the operation $1GSCO_x$ is associative over the words.
\end{lem}
\begin{proof}
\begin{equation*}
w_1>_1^x\rs w_2=\p_x(w_1)\cd x\cd\s_x(w_2).
\end{equation*}
\begin{eqnarray}\label{m5}
(w_1>_1^x\rs w_2)>_1^x\rs w_3 &=& [\p_x(w_1)\cd x\cd\s_x(w_2)]>_1^x\rs w_3\nonumber\\
    &=& \p_x[\p_x(w_1)\cd x\cd\s_x(w_2)]\cd x\cd \s_x(w_3)\nonumber \\
    &=& [(\p_x(\p_x(w_1)\cdot x))\cup
    (\p_x(w_1)\cd x\cd\p_x(\s_x(w_2)))]\cd x\cd \s_x(w_3)\nonumber\\
    &=& [\p_x(w_1)\cup \p_x(w_1)\cd x\cd\p_x(\s_x(w_2))]\cd x\cd \s_x(w_3)\nonumber\\
    && \hspace{1in}[\mbox{form definition it follows that $\p_x(\p_x(w)\cd x)=\p_x(w)$.}]\nonumber\\
    &=& \p_x(w_1)\cd x\cd \s_x(w_3)\cup\p_x(w_1)\cd x\cd\p_x(\s_x(w_2))\cd x\cd \s_x(w_3)\nonumber\\
    &&
\end{eqnarray}
On the other hand, consider
\[w_2>_1^x\rs w_3=\p_x(w_2)\cd x\cd \s_x(w_3).\]
\begin{eqnarray}\label{m6}
w_1>_1^x\rs (w_2>_1^x\rs w_3) &=& \p_x(w_1)\cd x\cd\s_x[\p_x(w_2)\cd x\cd \s_x(w_3)]\nonumber\\
    &=& \p_x(w_1)\cd x\cd[\s_x(\p_x(w_2))\cd x\cd \s_x(w_3)\cup(\s_x(x\cdot\s_x(w_3)))]\nonumber \\
    &=& \p_x(w_1)\cd x\cd \s_x(\p_x(w_2))\cd x\cd\s_x(w_3)\cup\p_x(w_1)\cd x\cd \s_x(w_3)\nonumber\\
    && \hspace{1in}[\mbox{form definition it follows that $\s_x(x\cd\s_x(w))=\s_x(w)$.}]\nonumber\\
    &=& \p_x(w_1)\cd x\cd \p_x(\s_x(w_2))\cd x\cd\s_x(w_3)\cup\p_x(w_1)\cd x\cd \s_x(w_3)\nonumber \\
    &&\hspace{1in}\mbox{from the previous lemma}
\end{eqnarray}
From (\ref{m5}) and (\ref{m6}), we have
\[(w_1>_1^x\rs w_2)>_1^x\rs w_3=w_1>_1^x\rs (w_2>_1^x\rs w_3). \]
\end{proof}
\begin{note}
Because of associativity of the operation we can write
\[ (w_1>_1^x\rs w_2)>_1^x\rs w_3=w_1>_1^x\rs (w_2>_1^x\rs w_3)=w_1>_1^x\rs w_2>_1^x\rs w_3.\]
\end{note}
\begin{corollary}
For any languages $L_1,~L_2,~L_3$ we can write
\[L_1>_1^x\rs(L_2>_1^x\rs L_3)=(L_1>_1^x\rs L_2)>_1^x\rs L_3.\]
\end{corollary}
\begin{proof}
The result is obvious as
\[L_1>_1^x\rs L_2=\bigcup_{\substack{w_1\in L_1\\w_2\in L_2}} (w_1>_1^x\rs w_2).\]
\end{proof}
\begin{lem}\label{as}
For any word $w\in GSCO_x^i(L)$, there exists a sequence of words
$w^0,w^1,\cdots,w^s\in L$ with $s\leq 2^i-1$, such that
\[w\in w^0>^x\rs w^1>^x\rs\cdots>^x\rs w^n.\]
\end{lem}
\begin{proof}
Let $w\in GSCO_x^i(L)$. We apply induction on $i$.
\par For $i=1;~w\in GSCO_x^1(L)$, i.e. $w\in GSCO_x(L,L)$. Hence, there exists
two words $w_0,w_1\in L $ such that \newline$w\in w_0>_1^x\rs w_1$. Note that
$w\in w>_1^x\rs w$ for any $w$. Hence if  $w\in L$, we shall write $w\in w>^x\rs w$.\\
\par Let the statement be true for each $i=1,2,\cdots, n$. We want to show
that it holds for $i=n+1$ as well.\\

Let $w\in GSCO^{n+1}(L)$. So there exists $w'$ and $w''\in
GSCO^n(L)$ such that $w\in w'>_1^x\rs w''$. By induction hypothesis,
we can express
\begin{gather*}
w'\in w_0'>^x\rs w_1'>^x\rs\cdots>^x\rs w_{2^n-1}'\hspace{1in}w_0',w_1',\cdots w_{2^n-1}'\in L\\
w''\in w_0''>^x\rs w_1''>^x\rs\cdots>^x\rs
w_{2^n-1}''\hspace{1in}w_0'',w_1'',\cdots w_{2^n-1}''\in L.
\end{gather*}
Hence
\[ w\in(w_0'>^x\rs w_1'>^x\rs\cdots>^x\rs w_{2^n-1}')>^x\rs(w_0''>^x\rs w_1''>^x\rs\cdots>^x\rs w_{2^n-1}'').\]
By associativity we can write
\[ w\in w_0'>^x\rs w_1'>^x\rs\cdots>^x\rs w_{2^n-1}'>^x\rs w_0''>^x\rs w_1''>^x\rs\cdots>^x\rs w_{2^n-1}''.\]
So $w$ can be generated by $x$-\sa of $2^{n+1}$ words  (may not be distinct) of $L$. Hence the lemma holds.
\end{proof}
\begin{theorem}\label{tt}
For any three words $w_1,~w_2,~w_3$ over $\sg^*$,
\[\bigcup_{a,b\in\sg}\big(w_1>^a\rs(w_2>^b\rs w_3)\big)=\bigcup_{a,b\in\sg}\big((w_1>^a\rs w_2)>^b\rs w_3\big).\]
\end{theorem}
\begin{proof}
Using the $\p_x$ and $\s_x$ notations mentioned earlier, we can write;
\begin{eqnarray*}
w_2>^b\rs w_3&=&\p_b(w_2)\cd b\cd \s_b(w_3),\\
w_1>^a\rs w_2&=&\p_a(w_1)\cd a\cd \s_a(w_2).
\end{eqnarray*}
\begin{eqnarray}\label{as1}
w_1>^a\rs(w_2>^b\rs w_3)&=&w_1>^a\rs(\p_b(w_2)\cd b\cd \s_b(w_3))\nonumber\\
        &=&\p_a(w_1)\cd a\cd\s_a\p_b(w_2)\cd b\cd\s_b(w_3) \nonumber\\      &&\cup~ \p_a(w_1)\cd a\cd\s_a\s_b(w_3).
\end{eqnarray}
Similarly we get
\begin{eqnarray}\label{as2}
(w_1>^a\rs w_2)>^b\rs w_3&=&(\p_a(w_1)\cd a\s_a(w_2))>^b\rs w_3\nonumber\\
        &=&\p_a(w_1)\cd a\cd\p_b\s_a(w_2)\cd b\cd\s_b(w_3)\nonumber\\
        &&\cup~\p_b\p_a(w_1)\cd b\cd\s_b(w_3).
\end{eqnarray}
The statement of the theorem can be restated as
\begin{multline*}
\big(\bigcup_{a=b}w_1>^a\rs(w_2>^b\rs w_3)\big)\bigcup\big( \bigcup_{a\neq b}w_1>^a\rs(w_2>^b\rs w_3)\big)=\\\big(\bigcup_{a=b}(w_1>^a\rs w_2)>^b\rs w_3\big)\bigcup\big(\bigcup_{a\neq b}(w_1>^a\rs w_2)>^b\rs w_3\big).
\end{multline*}
That is to prove the theorem, it is enough if we prove
\begin{multline}\label{as3}
\big(\bigcup_{a=b}w_1>^a\rs(w_2>^b\rs w_3)\big)\bigcup\big( \bigcup_{a\neq b}(\p_a(w_1)\cd a\cd\s_a\p_b(w_2)\cd b\cd\s_b(w_3)\\
\bigcup\p_a(w_1)\cd a\cd\s_a\s_b(w_3))\big)=\big(\bigcup_{a=b}(w_1>^a\rs w_2)>^b\rs w_3)\big)\bigcup\big( \bigcup_{a\neq b}(\p_b\p_a(w_1)\cd b\cd\s_b(w_3)
\\\bigcup\p_a(w_1)\cd a \cd\p_b\s_a(w_2)\cd b\cd\s_b(w_3))\big)
\end{multline}
Let
\begin{eqnarray*}
A &=&\bigcup_{a=b}w_1>^a\rs (w_2>^b\rs w_3)\\
    &=&\bigcup_a\{\p_a(w_1)\cd a\cd\s_a(w_3)\cup \p_a(w_1)\cd a\cd\s_a\p_a(w_2)\cd a\cd \s_a(w_3)\}
\end{eqnarray*}
\begin{eqnarray*}
C &=&\bigcup_{a=b}(w_1>^a\rs w_2)>^b\rs w_3\\
    &=&\bigcup_a\{\p_a(w_1)\cd a\cd\s_a(w_3)\cup \p_a(w_1)\cd a\cd\p_a\s_a(w_2)\cd a\cd \s_a(w_3)\}
\end{eqnarray*}
Using equation \ref{as1}, we define
\begin{eqnarray*}
B_1&=&\bigcup_{a\neq  b}\p_a(w_1)\cd a\cd\s_a\p_b(w_2)\cd b\cd\s_b(w_3)\\
B_2&=&\bigcup_{a\neq  b} \p_a(w_1)\cd a\cd\s_a\s_b(w_3)
\end{eqnarray*}
Using equation \ref{as2}, we define
\begin{eqnarray*}
D_1&=&\bigcup_{a\neq  b}\p_a(w_1)\cd a\cd\p_b\s_a(w_2)\cd b\cd\s_b(w_3)\\
D_2&=&\bigcup_{a\neq  b}\p_b\p_a(w_1)\cd b\cd\s_b(w_3)
\end{eqnarray*}
So from equation \ref{as3} it is sufficient to prove that
\begin{equation}\label{as4}
A\cup B_1\cup B_2=C\cup D_1\cup D_2.
\end{equation}
We p0rove next two lemmas which are required to prove equation
\ref{as4}.
\begin{lem}\label{asl1}
$B_2\subset A;~D_2\subset C$.
\end{lem}
\begin{proof}
\par We claim; given a word $w$, $\s_a\s_b(w)\subseteq\s_a(w)$.\\
\par Let
\begin{eqnarray*}
&& u\in\s_a(\s_b(w))\\
&\Rightarrow&\exists u'\in\sg^* \mbox{~such that~}u'au\in\s_b(w)\\
&\Rightarrow&\exists u''\in\sg^* \mbox{~such that~}u''bu'au=w\\
&\Rightarrow&w=(u''bu')au\\
&\Rightarrow&u\in\s_a(w).
\end{eqnarray*}
Similarly we can also prove that $\p_a\p_b(w)\subseteq\p_a(w)$. Note
that the other way is not true in general. Therefore
\begin{eqnarray*}
 \p_a(w_1)\cd a\cd \s_a\s_b(w_3)&\subseteq&\p_a(w_1)\cd a\cd\s_a(w_3)\\
        &\subset&\p_a(w_1)\cd a\cd\s_a(w_3)\\
        &&\cup\p_a(w_1)\cd a\cd\s_a\p_a(w_2)\cd a\cd \s_a(w_3).
\end{eqnarray*}
Taking union on both sides over $a\neq b$ we get $B_2\subset A$.\\
Similarly, we can prove that $D_2\subset C$. Hence the proof of
lemma.
\end{proof}
 In Lemma \ref{its2}, replacing $x$ by single symbol $a$, we get
\begin{equation}\label{as5}
A=C.
\end{equation}
By  Lemma \ref{asl1}, we have $B_2\subset A,$ and $D_2\subset C$. Hence from equation \ref{as4}, it is sufficient to prove that
\[A\cup B_1=C\cup D_1.\]

\begin{lem}\label{asl2}
For any word $w$,
\[\p_b(\s_a(w))=\s_a(\p_b(w)).\]
\end{lem}
\begin{proof}
The proof follows the same line of argument as of lemma \ref{its1}.
Let
\begin{eqnarray*}
&&u\in\p_b(\s_a(w)).\\
\Leftrightarrow&&\exists u'\in \sg^*~\mbox{such that~}ubu'\in\s_a(w)\\
\Leftrightarrow&&\exists u''\in \sg^*~\mbox{such that~}u''buau'=w\\
\Leftrightarrow&&u''au\in\p_b(w)\\
\Leftrightarrow&&u\in\s_a(\p_b(w)).
\end{eqnarray*}
Hence the proof.
\end{proof}
 Using the Lemma 6 it is obvious that
\begin{equation}\label{as6}
B_1=D_1.
\end{equation}
Combining equations \ref{as5} and \ref{as6}, we get our required result.
\end{proof}
\begin{corollary}
For any three words $w_1,w_2,w_3\in \sg^*$,
\[w_1>_1\rs(w_2>_1\rs w_3)=(w_1>_1\rs w_2)>_1\rs w_3.\]

\end{corollary}
\begin{proof}
By the corollary \ref{c1}, we have
\begin{eqnarray*}
w_1>_1\rs(w_2>_1\rs w_3)&=&w_1>_1\rs(\bigcup_{a\in\sg}w_2>_1^a\rs w_3)\\
        &=&\bigcup_{b\in\sg}w_1>_1^b\rs(\bigcup_{a\in\sg}w_2>_1^a\rs w_3)\\
        &=&\bigcup_{a,b\in\sg}w_1>_1^b\rs(w_2>_1^a\rs w_3).
\end{eqnarray*}
Similarly from the right hand side we get
\[ (w_1>_1\rs w_2)>_1\rs w_3=\bigcup_{a,b\in\sg}(w_1>_1^b\rs w_2)>_1^a\rs w_3.\]
Using the previous theorem, we have the required equality.
\end{proof}
\begin{theorem}
Any word $w\in1GSCO_x^i(L)$ can be written as
\(w\in1GSCO_x(1GSCO_x^j(L),L)\) for some $j$.
\end{theorem}
\begin{proof}
Let $w\in1GSCO_x^i(L)$. By the lemma \ref{as} we get a sequence of
words $w_1,w_2,\cdots,w_{n+1}\in L$, such that $w$ can be written in
the form
\begin{eqnarray*}
&&w\in w_1>^x\rs w_2>^x\ers\cdots w_n>^x\rs w_{n+1}\\
&&w\in (w_1>^x\rs w_2>^x\rs\cdots w_n)>^x\rs w_{n+1}\hspace{1in}\mbox{by associativity}\\
&\therefore&w\in1GSCO_x(w_1>^x\rs\cdots>^x\rs w_n,w_{n+1})\\
&\Rightarrow&w\in1GSCO_x(1GSCO_x^n(L),L).
\end{eqnarray*}
Hence the theorem.
\end{proof}
\par The above theorem suggests that the GSCO closure of $L$ can be proposed in another form, which we can call the restricted GSCO closure of $L$.
\begin{defin}
The restricted closure of GSCO denoted by $rGSCO^*(L)$ is defined
recursively as follows:-
\begin{eqnarray*}
rGSCO^0(L)&=&L\\
rGSCO^{i+1}(L)&=&rGSCO(rGSCO^i(L),L)\hspace{0.5in}i\geq1\\
rGSCO^*(L)&=&\bigcup_{i\geq0}rGSCO^i(L)
\end{eqnarray*}
\end{defin}
\par The main difference between $uGSCO^*$ and the $rGSCO^*$ is that, in the later case, crossover takes place between a word produced so far by the \sa and a words which is in $L$. In the former case, the crossover takes place between any pair of words generated so far. Interestingly the following theorem tells us that, they generate the same language.
\begin{theorem}
$r1GSCO^*(L)=u1GSCO^*(L)$.
\end{theorem}
\begin{proof}
By  definition it follows that
\[r1GSCO^*(L)\subseteq u1GSCO^*(L).\]
Hence it is enough if we show that
\[u1GSCO^*(L)\subseteq r1GSCO^*(L).\]
Let $w\in u1GSCO^*(L)$. Hence $w\in uGSCO^i(L)$ for some $i$. Hence
there exists $w_0,w_1,\cdots,w_{2^i-1}\in L$ such that
\begin{eqnarray*}
&&w\in\big((\cdots((w_0\ers w_1)\ers(w_2\ers w_3))\ers\cdots\ers(w_{2^i-2}\ers w_{2^i-1}))\cdots)\big)\\
&\Rightarrow&w\in w_0\ers w_1\ers\cdots\ers w_{2^i-1}\hspace{1in}\mbox{Since $\ers$ is associative}\\
&\Rightarrow&w\in \big(\cdots((w_0\ers w_1)\ers w_2)\ers\cdots w_{2^n-3})\ers w_{2^n-2}\big)\ers w_{2^n-1}\\
&\Rightarrow&w\in GSCO\big(\cdots GSCO(\cdots( GSCO(GSCO(GSCO(w_0,w_1),w_2),w_3)\cdots),w_{2^i-2}),w_{2^i-1}\big)\\
&\Rightarrow&w\in GSCO(\cdots(GSCO(GSCO(L),L)\cdots),L)\\
&\Rightarrow&w\in w\in r1GSCO^{2^i-1}(L).
\end{eqnarray*}
Hence the theorem.
\end{proof}
\begin{note}
We can also prove the above theorem by using closure property of
$r1GSCO^*(L)$ under the 1-GSCO.
\end{note}
\par Because of this theorem, we no more distinguish $rGSCO^*(L)$ and $uGSCO^*(L)$ and we simply refer them as $GSCO^*(L)$. The proof also shows that we can construct $GSCO^*(L)$ as follows
\[GSCO^{n+1}(L)=GSCO(L,GSCO^n(L)).\]
\section{Regularity of GSCO}
\begin{defin}[Base of a word]
Base of a word $w$, denoted by $B(w)$ is the minimal set of words
whose iterated \sa generates $w$ in a way that every element of
$B(w)$ takes part in GSCO at least once.
\[B(w)=\{u_1,u_2,\cdots,u_k:w\in GSCO^*(\{u_1,u_2,\cdots,u_k\})\}.\]
\par Here the word `minimal' is used in the sense that if there exists
$B'(w)= \{u_1',u_2',\cdots,u_{k}'$, $w\in GSCO^*(\{u_1',u_2',$
$\cdots,u_k'\})\}$ such that $B'(w)\subseteq B(w)$, then
$B(w)=B'(w)$.
\end{defin}
\par $B(w)$ is the set of minimal words to generate $w$ by the process of GSCO.
$B(w)$ is a finite set for any word $w$. $B(w)$ need not be unique for a word $w$.
For an example $B(abbbc)=\{ab,bb,bc\}$ and $B(abbbc)=\{abb,bc\}$. $B(w)$ will be called
$nB(w)$ if all the words of $B(w)$ are of length $n$. $nB(w)$ with $n>2$ is not unique.
 As an example, the word $w=abbbc$ has two $4B$ sets which are
  $4B(abbbc)=\{abbb,bbbc\}$ as well as $\{abba,bbbc\}$.
  For words $w$ such
 that $|w|=1, w\in1B(w)$. It is interesting to note that $2B(w)$ is unique for a word.
  For $w=a_1a_2\cdots a_k$, $a_i\in\sg,~i=1,2,\cdots, k$
\[
2B(w)=\{a_1a_2,a_2a_3,\cdots,a_{k-1}a_k\}.\]
 For $a~\in \Sigma~,
2B(a)$ is taken as  the set $\{a\}$ and $2B(\ve)~=~\ve$. We define
$2B$ of a language $L$ as $2B(L)~=~\cup_{w \in L} 2B(w)$. For
example, $2B(a^+)~=~\{a, aa\}$.
\begin{theorem}
For a language $L$, $GSCO^*(L)$  is a regular language.
\end{theorem}
\begin{proof}
Let $ \Sigma $ be the alphabet of $L$.  We define a relation $R$
over $ \Sigma^*~\times~\Sigma^*$ such that \[ x R y~~ {\text iff~~}
2B(x)~=~2B(y); ~~\sg_{1} (x)~=~\sg_{1}(y);~~
\sg_{|x|}(x)~=~\sg_{|y|}(y). \symbolfootnote[1]{$\sg_i(w)$ is the
symbol in the $i^{th}$ position of the word $w$.}
\]
\\
\noindent{\bf Claim 1 :} $R$ is a right invariant (with respect to
concatenation) equivalence relation.\\

 $R$ is reflexive, since $xRx$. $R$ is symmetric since
$xRy\Rightarrow yRx$. If $xRy$ and $yRz$, we have $2B(x)=2B(y)~;~
\sg_{1} (x)~=~\sg_{1}(y)$ $ and $ $\sg_{|x|}(x)=\sg_{|y|}(y)$ and
$2B(y)=2B(z); \sg_{1} (y)~=~\sg_{1}(z); and
\sg_{|y|}(y)=\sg_{|z|}(z)$. Hence, we have  $2B(x)=2B(z); \sg_{1}
(x)~=~\sg_{1}(z); \sg_{|x|}(x)=\sg_{|z|}(z)$ implies the
transitivity of $R$. Hence $R$ is an equivalence relation.

\par Let $xRy$. So
\begin{equation} \label{bs}
2B(x)=2B(y);\hspace{1cm}\sg_{1}(x)=\sg_{1}(y);\hspace{1cm}
~~\sg_{|x|}(x)=\sg_{|y|}(y).
\end{equation}

Let  $z$ be any word.
\[2B(xz)=2B(x)\cup\{\sg_{|x|}(x)\cd\sg_1(z)\}\cup2B(z).\]
Similarly
\[2B(yz)=2B(y)\cup\{\sg_{|y|}(y)\cd\sg_1(z)\}\cup2B(z).\]
By \ref{bs} we have
\[2B(xz)=2B(yz)~; \hspace{1cm} \sg_{1}(xz)=\sg_{1}(yz); \hspace{1cm} ~~~~~~~~
~\sg_{|x|+|z|}(xz)=\sg_{|y|+|z|}(yz),\] which implies $xzRyz$. Hence
$R$ is a right invariant with respect
to concatenation.\\\\

\noindent{\bf Claim 2 :} Number of equivalence classes of $R$ over $
\Sigma^*$ is finite.

\par Every equivalence classes of $\Sigma^*$ will have a $2B$ set, a
symbol $s\in \Sigma$ and a symbol $e\in \Sigma$ such that the
elements in the equivalence class are just the elements of
$GSCO^*(2B) \cap s.\Sigma^*.e$.  Every equivalence class is
parameterized by a $2B$ set, a symbol $s$(which is  the starting
symbol of the words in that equivalence class) and the symbol
$e$(which is the ending symbol of the words in that class).  We
denote an equivalence class by $\langle s,2B,e\rangle,  ~ 2B \in
  2^{\Sigma ^2} \backslash  \emptyset , ~ s,e  \in   \Sigma.$  For example,
if $\Sigma = \{a,b\}$, abbbb will be in the equivalence class
$\langle a,\{ab,bb\},b \rangle$. The words $w \in \Sigma^*$ such
that $|w| = 1$, will be related to itself under the relation $R$ and
not to any other words other than $\Sigma^* $. That is, these words
will be in the equivalence class in which only one word  $w$ will be
present. The word `$a \in \sg$' will be present in one equivalence
and no other element will be present in that equivalence class.
Similarly, the element `$b$' will be present in one equivalence
class.  We denote the equivalence classes which has only one element
of length one by $\langle a, \{a\},a \rangle,   a \in \Sigma$.  The
word $\ve \in \Sigma^*$ will be in an equivalence class which will
not have any other element of $\Sigma^*$ in it. Thus we have two
categories
of equivalence classes.\\
 $ Category ~  I  ~~ : ~~  \langle s,2B,e \rangle,  ~ 2B \in
  2^{\Sigma ^2}
\backslash  \emptyset ,~  s,e  \in   \Sigma.\\
Category  ~ II ~ ~:  ~~ \langle a, \{a\},a \rangle,  ~ a \in
\Sigma$.
\\
For every equivalence class of Category I, we have the triple
$\langle s,2B,e \rangle ,~ 2B \in   2^{\Sigma ^2}\backslash
\emptyset , ~s,e  \in \Sigma$. For every triple $\langle  s,2B,e
\rangle ,~ 2B \in 2^{\Sigma ^2}\backslash \emptyset , ~ s,e  \in
\Sigma$, we have an equivalence class of $R$ ( some equivalence
classes of $R$ over $\Sigma^*$ may be empty).  That is, the triple
$\langle s,2B,e\rangle $ characterizes an equivalence class of $R$.
If $|\Sigma| = n,~ | 2^{\Sigma ^2}\backslash  \emptyset  |  =
2^{n^2}-1$. The number of such triples will be $(2^{n^2}-1) \times
n^2$.  That is under category I, the total number of equivalence
classes of $R$ over $\Sigma^*$ will be $n^2(2^{n^2}-1)$.  Under
category II, the number of equivalence classes will be the number of
triples of the form $\langle a, \{a\},a \rangle , a \in \Sigma,   a
\in
 \Sigma \cup \{\varepsilon \}$.  Under category II, the total number
of equivalence classes are $n+1$. The total number of equivalence
classes of $R$ will be $
n^2(2^{n^2}-1) + (n+1)$, which is finite since $n$ is finite.\\\\
{\bf Claim 3 : ~~}$ GSCO^*(L)$ is the union of some of the
equivalence classes of $R$.

\par Since $ GSCO^*(L) \subset  \Sigma^*,$ the elements of $GSCO^*(L)$
will be spread out in different equivalence classes of $R$ over
$\Sigma^*$.  $\ve \notin GSCO^*(L)$.  If the symbol $a \in
 \Sigma$ such that   $a \in GSCO^*(L)$, then $a$  will be present in the
equivalence class $\langle a, \{a\},a \rangle $ and no other element
other than `$a$' will be present in $\langle a, \{a\},a \rangle $.
So the equivalent classes of category II will be contained in
$GSC0^*(L)$ if $a\in GSC0^*(L)$.\\

We prove the following claim to show  that, if there is an
equivalence class of category I which shares at least one common
word with $GSCO^*(L)$, then that equivalence class will be fully
contained in $GSCO^*(L)$.\\
\\
\noindent {\bf Claim 3(a):  } If $GSCO^*(L) \cap \langle
s,2B,e\rangle \neq \emptyset  $,for some $s, e, 2B,  $ then $\langle
s, 2B, e\rangle \subseteq GSCO^*(L).  $
\par  We have to prove $\langle s, 2B,
e\rangle \subseteq GSCO^*(L).  $ Suppose the other way. That is,
there exists a word $w$ such that $|w| >  1 $, $w \in \langle s, 2B,
e\rangle  $ and $ w \notin GSCO^*(L)$. Since $w \in \langle s, 2B,
e\rangle  $ we have  $w \in GSCO^*(2B) \cap s.\Sigma^*.e$.  Let $w =
a_1a_2 \cdots a_{n}, |w|>1.$ Here $s = a_1$; $e = a_n$. $ w \in
a_1a_2\ers a_2a_3\ers \cdots
 \ers a_{n-1}a_{n},  ~~~ a_ia_{i+1} \in 2B, ~~~ i =
 1,2,3\cdots,n-1$. We want to show that there exists a sequence of
 words in $GSCO^*(L)$, which by iterative crossover can generate $w$. The
 following claim helps us to get such a sequence of words.
\\

\noindent{\bf Claim 3(b):  }
\begin{enumerate}
\item There
exists words $w_i = u_ia_ia_{i+1}v_i \in GSCO^*(L)$, for some $u_i
,v_i \in
 \Sigma^*$.
 \item There exists a word $w_1 \in  GSCO^*(L)$ such
that $a_1a_2 \in \p(w_1) $.
\item There exists a word $w_{n-1} \in
GSCO^*(L)$ such that $a_{n-1}a_n \in \s(w_{n-1})$.

\end{enumerate}
\par Elements of $2B $ (which is under consideration in Claim 3(a)) are
in $2B(GSCO^*(L))$.  That is, there exists a word of the form
$ua_1a_2v \in  GSCO^*(L)$.
\par
Since the first symbol of $w$ is $a_1$, $s = a_1$, there exists a
word $a_1t \in GSCO^*(L), ~t \in \Sigma^*$. Since $GSCO^*(L)$ is a
crossover language and $a_1t,ua_1a_2v \in GSCO^*(L),$ $a_1a_2v \in
a_1t \ers ua_1a_2v \in GSCO^*(L)$.  We write $w_1 = a_1a_2v \in
GSCO^*(L)$. Similarly, there exists a word $w_{n-1} = v'a_{n-1}a_n
\in GSCO^*(L)$, for some $v' \in \Sigma^*$.

\par The set $2B$ (which is under consideration) contains all the sub
words of length 2 of some words in $GSCO^*(L)$. (that is, the set
$2B$ contains all the sub words of length 2 for the words which are
present in the equivalence class $\langle s,2B,e\rangle $.  For each
$a_ia_{i+1} \in  2B,  i = 1,2,3 \cdots, (n-2)$, there exists a word
$w_i = u_ia_ia_{i+1}v_i \in GSCO^*(L)$, for some $u_i,v_i \in
\Sigma^*$ ($w_i's$ need not be distinct). Thus we have a sequence of
words $w_i \in GSCO^*(L)$. Thus we have the claim 3(b).

\par Clearly $a_1a_2a_3\cdots a_n\in a_1a_2v \ers u_2a_2a_3\cdots\ers
v'a_{n-1}a_n$. That is, $w \in
 w_1 \ers w_2 \ers \cdots w_n,  w_i \in GSCO^*(L)$. Thus,
$w \in GSCO^*(L)$, which contradicts with the assumption that $w
\notin GSCO^*(L)$.  Hence, we have the claim 3(a).

\par Thus, we have , for every $a \in  GSCO^*(L)$ (such that $a \in
\Sigma)$, the equivalence class of category II which contains $a$,
viz.,$ \langle a,\{a\},a\rangle $ will be fully contained in
$GSCO^*(L)$ since $\langle a,\{a\},a\rangle $ contains only one
element $a$.  For every $w \in GSCO^*(L)$,  $|w| > 1$, the
equivalence class (of category I) which contains $w$, viz., $\langle
s,2B,e\rangle $ will be fully contained in $GSCO^*(L)$. Thus,
\[
GSCO^*(L) =  (\bigcup_{a \in GSCO^*(L) \cap \Sigma} \langle
a,\{a\},a\rangle ) \bigcup (\bigcup_{w \in GSCO^*(L) \cap \langle
s,2B,e\rangle } \langle s,2B,e\rangle). \]

 We know that $R$ is of finite index.  Hence, $GSCO^*(L)$ is the
 union of some of the equivalence classes of a right invariant
 equivalence relation of finite index. Thus, by  Myhill - Nerode theorem,
 $GSCO^*(L)$ is regular.
\end{proof}
\par The converse of this theorem is not true, i.e. not all regular
 language can be obtained by using GSCO. We give a counter example in example \ref{not}.

\begin{defin}
A language $L$ is said to be a crossover language if there exist a
set $L'$ such that $GSCO^*(L') = L$.  That is, $L$ is called an
crossover language if $L$ can be got by the iterated GSCO process of
some set $L'$.
\end{defin}
\begin{example}
1. $L = \{a,b\}$ is a crossover language  $GSCO^*(L) = \{a,b\}$.\\
 2. $L = a^+b^+$ is a crossover language since $GSCO^*(\{aabb,
aaabbb\})  =   a^+b^+$
\end{example}
\begin{rem}
All crossover languages are regular and no crossover language will
contain $\ve$ (word of length 0)
\end{rem}

\begin{theorem}
A language $L$ is said to be a crossover language if and only if $L$
is closed with respect to the operation $GSCO$.
\end{theorem}
\begin{proof}
Given $L$ is a crossover language. Then there exist a language $L'$
such that $GSCO^*(L') = L$  Let $x,y \in L$. Then $x,y \in
GSCO^*(L')$. $GSCO(x,y)  \in  GSCO^*(L')$ since $GSCO^*(L')$
 is the transitive closure of GSCO.   Hence, $GSCO(x,y) \in L$, since
 $GSCO^*(L') = L$.\\

 \noindent The other way proof :\\

 \par Suppose $L$ is closed with respect to GSCO. $GSCO(x,y) \in
 L,  $ for every $ x,y \in L$.  $GSCO(GSCO(x,y),z) \in L, \forall x,y,z \in
  L$. That is, $GSCO^2(L) \subseteq L$.  Continuing like this, we have
 $GSCO^i(L) \subseteq L, i \geq 0$. Then, $\cup_{i}GSCO^i(L) \subseteq L$,
 and
   $L \subseteq \cup_{i}GSCO^i(L)$. Hence $GSCO^*(L) = L$ which implies
 that $L$ is a crossover language.
\end{proof}
\begin{example}\label{not}
The language $L=\{a^{2n}:n\geq1\}$ is a regular language. However it
is not a GSCO language as it is not closed under GSCO operation.
$a^3\in a^2\ers a^2$ but $a^3\notin L$.
\end{example}
\begin{theorem}
For any crossover language $L$, there exists  three finite sets,
$S,E
 \subseteq \Sigma  , B \subseteq \Sigma^2$ such that
\[L = (GSCO^*(B)  \cap S \Sigma^* E)  \bigcup  (L \cap \Sigma) \]
\end{theorem}
\begin{proof}
Given a crossover language $L$, $L$ will not contain $\ve$. Since
$L$ is regular, we can find a right-linear grammar $G =(N,T,P,S)$
such that $G$ generates $L$. Without loss of any generality, let $G$
be a grammar without $\ve$ - productions (since $L$ does not have
$\ve$), unit productions and any useless symbols. We construct a set
$B$ (called the {\em Base set} of $L$) as follows.
\begin{enumerate}
\item For every production $ S \rightarrow  a  \in P,   a \in T $; include $a \in
B$.
\item For every pair of  productions $ X\rightarrow aA,~    A \rightarrow bB  \in
P;
 ~a , b\in T;~
A,B,X \in N$; include $ab \in B$.
\item For every pair of  productions $ X\rightarrow aA,~  A \rightarrow b  \in P; ~ a, b \in
T;~ A,X \in N$; include $ab \in B$.
\end{enumerate}
The construction of $B$ tells that the set $B$ contains all the sub
words of length 2 of $L$. We construct the set $S$(Start symbol set)
and $E$(end symbol set)as follows.
\begin{enumerate}
  \item    For a production $ S \rightarrow a,$ include $a \in  S$

\item For a production $ S \rightarrow aA,$ include $a \in  S$

\item For a production $ A \rightarrow a,$ include $a \in  E$
\end{enumerate}
$S$ and $E$ will have the first and the last symbol of the words of
$L$.\\

 \noindent {\bf Case I:} All the words in $L$ of length greater than or equal to 2 are
 in $GSCO^*(B)  \cap S \Sigma^* E$ and vice-versa.

\noindent {\bf Part I :} Let $w  =  a_1a_2 \ldots a_n \in L,~ |w|
\geq 2$. Then $w \in a_1a_2 >_1\rs a_2a_3 >_1\rs  \ldots >_1\rs
a_{n-1}a_n$. Since $w \in L, ~a_1  \in  S$ and $a_n \in  E.$  Here,
$a_ia_{i+1}, i = 1,2, \ldots (n-1),$ are the sub words of $L$ of
length 2 .  This implies $a_ia_{i+1} \in B ,\forall i$. $w \in
GSCO^*(B)$ and $w \in S \Sigma^* E$. Thus, we have $w \in
 GSCO^*(B) \cap S \Sigma^* E$.\\

\noindent {\bf Part II:} Let $w \in  GSCO^*(B) \cap S \Sigma^* E$.
Let $w = a_1a_2 \ldots a_n, a_1 \in S $ and $a_n \in  E$. $w \in
a_1a_2
>_1\rs a_2a_3 >_1\rs  \ldots >_1\rs a_{n-1}a_n,~  a_ia_{i+1} \in B,~1\leq i\leq n-1$. We
claim now that there exists a word $w_1 \in L$ such that $a_1a_2 \in
Prefix(w_1)$. Here, $a_1a_2 \in B$. Since $B$ contains all the sub
words  of $L$ of length 2, there exists a word of the form $ua_1a_2v
\in L, u,v \in \Sigma^*$.  Since $a_1 \in S $, there exist a word
$a_1t \in L$. Further, $a_1t >_1\rs ua_1a_2v \subseteq L$. That is,
$a_1a_2u \in L$ Thus, we have the claim of the existence of $w_1 \in
L$ whose prefix is $a_1a_2$. Similarly, we can prove that there
exists a word $w_{n-1} =va_{n-1}a_n$in $L$. Since $B$ contains all
the sub words  of $L$ of length 2, for each
$a_ia_{i+1},`i=2,3,\ldots (n-2)$, there exists a word $w_i =
u_ia_ia_{i+1}v_i \in L.$  Thus, we have a sequence of words $w_i \in
L ,  i =1,2,...n$.  Clearly, $a_1 \ldots a_n  \in a_1a_2u >_1\rs
u_2a_2a_3v_2 >_1\rs \ldots va_{n-1}a_n$. That is, $ w \in  w_1
>_1\rs w_2 \ldots >_1\rs w_n ,~ w_1,w_2,\ldots w_n \in L$.  Since
$L$ is a crossover language, $w \in GSCO^*(L) \subseteq L$.  Hence,
$w \in L$.
\\

\noindent{\bf Case II:} All the words in $L$ of length equal to 1
are in $L\cap\sg$ and vice-versa.

  Since $GSCO^*(B) \cap S
\Sigma^* E$ contains only words in $L$ of length $\geq2$, it is
clear that the words  $w\in L$ of length 1 are in $ L \cap
\Sigma$.\\
Hence
\[L = (GSCO^*(B)  \cap S \Sigma^* E)  \bigcup  (L \cap \Sigma) \]

\begin{corollary}
If $L \cap \sg = \emptyset$ (that is, $L$ does not contain any word
of length 1), then $L = GSCO^*(B) \cap S \sg^* E $
\end{corollary}
Proof is immediate.
\begin{corollary} Let $\sg$ be an alphabet.
If all the words in $L$ are  of the form $\sg \sg^* \sg$, (that is,
if $L$ contains words which starts with all the possible symbols and
ends with all the symbols, then $L=GSCO^*(B)$
\end{corollary}
Proof is immediate since $S = \sg = E$ \\
Given a crossover language $L$, the above theorem gives the
construction of the set $B$ with which one can generate $L$ by the
iterative $GSCO$.  The base set of a  crossover language $L$ will
have all the sub words of $L$ of  length 2 along with the words of
length 1 in $L$ where as the $2B$ set of $L$ will contain all the
sub words of $L$ of length 2, words of $L$ of  length 1 and $\ve$
(word of length 0) if $\ve$ is in $L$. In other words, if $L$ does
not contain $\ve$, then the base set of $L$ and $2B(L)$ will be the
same. The next lemma shows that the base set of a crossover language
is unique.
\begin{lemma}
 The base set of a crossover language is unique.
\end{lemma}
Proof is obvious.

\end{proof}

\section{Comparison with other sub-regular families}
Since $GSCO^*(L)$ is a subclass of regular languages, in this
section, we compare the various subclasses of regular languages with
the crossover language. For this
 purpose, we consider different classes of crossover languages as follows.

\begin{defin}
We define the following classes of \sa languages based on $R$, the
set of overlapping. Let $\sg$ be the alphabet of the axiom.
\begin{description}

\item[TSyGSCO] Class of languages that can be generated  by the operation
$GSCO^*_R$ over an  axiom, where $R = \sg$ .
\item[SyGSCO] Class of languages that can be generated by the operation $GSCO^*_R$
over an  axiom, where $R \subseteq \sg$.
\item[StGSCO] Class of languages that can be generated by the operation  $GSCO^*_R$
over an  axiom, where $R \subseteq \sg^+$.
\item[TStGSCO] Class of languages that can be generated by the operation $GSCO_R$
over an axiom, where $R =\sg^+$.
\end{description}
\end{defin}
\begin{theorem}

\begin{enumerate}
\item $TSyGSCO\subset SyGSCO$
\item $SyGSCO\subset StGSCO$
\item $TStGSCO \subset StGSCO$
\item $TStGSCO=~TSyGSCO$
\end{enumerate}
\end{theorem}
\begin{proof}
Let $L~\in TSyGSCO$. Then there exists a set $R$ and a set $L_0$
such that $L=GSCO_R^*(L_0)$, where $R$ is the alphabet of $L_0$.
That is, $L$ is generated by the crossover where all the overlapping
are over the symbols of the alphabet of $L$. We have
$L=GSCO_R^*(L_0)$, $R \subseteq \mbox{~alphabet of~} L_0$, which
implies $L\in SyGSCO$. The other way is not true. The language
$a^+b^2\in SyGSCO$, because $a^+b^2~=~ GSCO_{a}(a^+b^2)$.  But,
$a^+b^2~ \notin TSyGSCO$, since the language $a^+b^2$ is not closed
w.r.t the operation $GSCO_b$.

\par Let $L\in SyGSCO$. Then there exist a set $R$ and a set $L_0$
such that $L=GSCO_R^*(L_0)$, $R\subset \Sigma_L$.  Since
$R\subset\Sigma_L$, $R\subset{\Sigma_L}^*$. This implies $L \in
StGSCO$. The other way is not true. $(aa)^+b^2(aa)^+\in SyGSCO$ with
respect to  $R=\{b^2\}$. However this language is not closed with
respect to any symbol.
\par  Let $L\in TStGSCO$. Then there exist an $L_0, R$ such
that
 $L=GSCO_R^\star{L_0}$, where $R~=sub(L_0)$.  Since $R \subset  \Sigma_{L_0}^*$,
  $L\in SyGSCO$. The other way is not true. The language $a^+b^2\in
  StGSCO_R$ where $R=\{b^2\}$, but it is not in $TStGSCO$.
\par Immediate from corollary 1.
\end{proof}
\par
\begin{figure}[h]
\setlength{\unitlength}{2mm}
\begin{center}
\begin{picture}(50,45)(-25,-20)
\put(-2,-20){$MON$} \put(-0.5,-18){\vector(-2,1){7}}
\put(-10,-14){$COMB$} \put(-9,-13){\vector(-2,1){13}}
\put(-24,-6){$DEF$} \put(-8,-12){\vector(0,1){3}}
\put(-10,-8){$GSCO$}
\put(-8,-6){\vector(0,1){3}}\put(-12,-2){$SH=SyGSCO$}
\put(-8,-1){\vector(0,1){3}}\put(-15,3){$SLT=NCH=StGSCO$}
\put(-8,4.5){\vector(0,1){8}}\put(-10,13){$ESF$}
\put(-8,14){\vector(0,1){3}}\put(-9,17){$PS$}
\put(-22,-4){\vector(2,3){12}}
\put(0,-18){\vector(0,1){5}}\put(-2,-12){$NIL$}
\put(0,-10){\vector(1,4){1}}\put(0,-6){$ORD$}
\qbezier(1,-5)(10,7)(-7,13)\put(-6,12){\vector(-1,1){1}}
\put(1,-18){\vector(2,1){8}}\put(6,-14){$COMM$}
\put(-8,19){\vector(4,1){8}}\put(0,22){$REG$}
\qbezier(8,-12)(6,10)(3,20)\put(3,20){\vector(0,1){1}}
\put(10,0){$SUF$}
\qbezier(12,1)(14,10)(4,21) \put(4,21){\vector(-3,1){1}}
\qbezier(2,-18)(20,-15.5)(12,-0.5)
\put(12.55,-1.5){\vector(-2,3){1}}
\end{picture}
\end{center}
\caption{Relations between different subclasses of the regular
language and their relations with $GSCO$} \label{rel}
\end{figure}
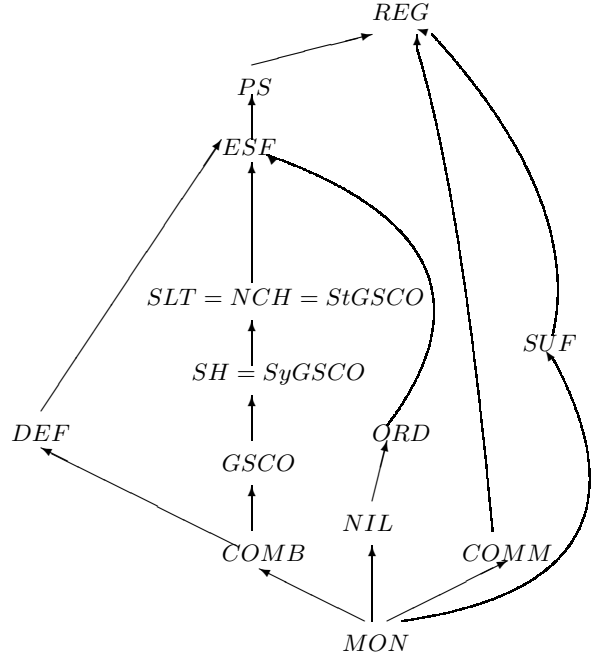

 We examine now the relationships of class of GSCO languages with
a series of well-known subfamilies of $REG$, considered  in \cite{Mc,swat,MPS}.
\begin{defin}
A language $L \subseteq \sg^* $is called
\begin{description}
\item[Combinational]  if and only if  $ L = \sg^*U$, for some $ U \subset \sg$;
\item [Definite]  if and only if  $ L=L_1 \cup \sg^*L_2$, where $L_1, L_2$ are finite subsets of  $\sg^*$;
\item [Nil-potent]  if and only if  either $L$ or $\sg^*-L$ is finite;
\item [Commutative]  if and only if  $x \in L$ implies that all permutations of $x$ are in $L$;
\item [Suffix-closed]  if and only if  $\s(L) \subset L)$
\item [Non-counting (extended star-free)]  if and only if  there is an integer $k\geq1$ such that
for every $x,y,z \in \sg^*, y \neq \ve$, we have $xy^kz \in L$
\item [Power-separating]  if and only if  for each $x\in \sg^*$ there is a natural number
$m \geq 1 $
 such
that either $L \cap \{ x^n|~n ~\geq~ m~\} = \emptyset$  or $ \{x^n
|n \geq m\} \subset L$.
\item [Ordered]  if and only if  $L$ is accepted by some deterministic finite
automaton $(K,\sg,\delta, s_0, F)$ with a totally ordered set of
states $K$, such that for each $a \in \sg$, the relation  $s \leq
s'$ implies $\delta(s,a) \leq \delta(s',a)$.
\end{description}
\par We denote by $COMB,DEF,NIL,COMM,SUF,ESF,PS,ORD$ the families of combinational,
 definite, nilpotent, commutative, suffix-closed, non-counting, power-separating and
 ordered languages. The relation between the different type of GSCO classes and
 the above sub classes are given in the figure \ref{rel}.
\end{defin}

\begin{theorem}
$SH = SyGSCO ;~~ NCH = StGSCO$.
\end{theorem}
\begin{proof} We have to show that a language $L$
which can be generated by a simple splicing system can also be
generated by $StGSCO$ and vice versa. For that, it is enough if we
show that, for any axiom $A$, there exists a language $L_0$ such
that
\[\sigma^*(A)=GSCO^*_R(L_0),\]
for some $R$ and vice versa.  We use  the method of induction.
Consider $ \sigma_R^*(A)$, $R$ is the subset of the alphabet of $A$
 where $ \sigma $ is the splicing scheme of a simple splicing
system.  Let $L_0~=A$. $x\in L$ implies $x \in \sigma_R^i(A),
i\geq0$.
 If $i=0$, $x\in A =L_0~.$  We assume that for an $i > 0,$, $y \in
  \sigma_R^i(A)$, we have that $y \in
GSCO_R^*(L_0)$. Let $w \in \sigma_R^{i+1}(A)$. Then, there exist
$w_1,w_2 \in \sigma_R^i(A)$ such that $(w_1, w_2)\vdash_a~ w, ~a\in
R$. That is, $w_1=u_1au_2, w_2=v_1av_2, w=u_1av_2$.
 By the process of induction, $u_1au_2, v_1av_2~\in GSCO_R^*(L_0)$. Then
  $GSCO_R(u_1au_2, v_1av_2) = u_1av_2=w \in GSCO_R^*(L_0)$
Hence, $\sigma^*(A) \subseteq GSCO^*_R(L_0)$.

  \par On the other hand, consider $GSCO_R^*(L_0)$
  where $R$ is a subset of the alphabet of $L_0$. Let $w\in GSCO_R^*(L_0)$. So $w\in
  GSCO_R^i(L_0)$ for some $i$. Let us put $L_0=A$.
\par If $i=0$, then $w\in\sigma_R^0(A)$ holds trivially. Using the
induction hypothesis, assume that any $y\in GSCO^i(L_0)$,
$y\in\sigma_R^*(A)$.
\par Let $w\in GSCO^{i+1}(L_0)$. So, $w\in GSCO(w_1,w_2)$, where
$w_1,w_ 2\in GSCO^i(L_0)$. So there exists an $a\in R$ such that
$w\in GSCO_R(w_1,w_2),~a\in R$ i.e. $w_1=u_1au_2,~w_2=v_1av_2$ and
$w=u_1av_2$.
\par By the process of induction, both $w_1\in\sigma_R^*(A)$ and $w_2\in\sigma_R^*(A)$
implies $w\in\sigma_R^*(A)$ since $(w_1,w_2)\vdash_a w$. Hence $
GSCO^*_R(L_0)  \subseteq  \sigma^*(A)$, which proves $ SH = SyGSCO$.
  Similarly, we can prove $NCH = StGSCO$.

\end{proof}
\begin{theorem}
$L\in TSyGSCO$ if and only if $L$ is closed with respect to the the
operation $GSCO_a~\forall a\in \sg_L$ (alphabet of $L$).
\end{theorem}
\begin{proof}
Let $L\in TStGSCO$. Hence there exists a set $L_0$ such that
$L=GSCO_R^*(L_0), ~R=\sg$. This implies, $L$ is closed with respect
to $GSCO_a, ~\forall a\in\sg_L$.
\par  Let $L$ is  closed with respect to the operation $GSCO_a, ~\forall
a\in\sg$, i.e. $GSCO_a(x,y)\in L~\forall x,y\in L$ and $\forall a\in
\sg$. This implies $GSCO_a^1(L)\subseteq L$. Now
$GSCO_a(GSCO_a(x,y),z)\in L~\forall x,y,z \in L$, i.e.
$GSCO_a^2(L)\subseteq L$.
\par Continuing on the same line, we get
\begin{eqnarray*}
 & GSCO_a^i(L)\subseteq L\\
 \Rightarrow & \bigcup_iGSCO_a^i(L)\subseteq L.
\end{eqnarray*}
Since we have $L\subseteq \bigcup_iGSCO^i(L)$,  we conclude
\[GSCO^*(L)=L,\]
i.e. $L$ is a $TStGSCO$.
\end{proof}
\begin{theorem}

$L\in StGSCO$ if and only if $\exists R\in \sg_L^*$ such that $L$ is
closed with respect to the operation $GSCO_R$.
\end{theorem}

\begin{proof}
$L\in StGSCO$, implies $\exists$ a set $L_0$ and $R$ such that
\[L=GSCO_R^*(L_0),~~R\subseteq \sg_L^*.\]
Hence $L$ is closed with respect to the operation $GSCO_R$.
\par For  $R\subseteq \sg_L^*$. Let  $L$
be closed with respect to the operation $GSCO_R$. So,
 \begin{eqnarray*}
  \forall x,y\in
L&GSCO_R(x,y)=z\in L.\\
  i.e. & GSCO(GSCO_R(x,y),z)\subseteq L, ~\forall x,y,z\in L\\
i.e. & GSCO^2(L)\subseteq L.
\end{eqnarray*}
Continuing on the same lines,
\begin{eqnarray*}
GSCO^i(L)&\subseteq& L,~\forall i\geq 0\\
\bigcup_i GSCO_R^i(L)&\subseteq& L\\
\Rightarrow ~GSCO_R^*(L)&\subseteq& L
\end{eqnarray*}
Hence, we can conclude that
\[L=GSCO_R^*(L).\]
\end{proof}
\par Head has proved that NCH=SLT \cite{head1}. Thus we have the
following theorem whose proof is immediate.
\begin{theorem}
$L\in SLT$ if and only if there exists $R\subseteq\sg_L^*$ such that
$L$ is closed with respect to the operation $GSCO_R$.
\end{theorem}

\begin{theorem}
$SLT \subset ESF$
\end{theorem}
\begin{proof}
Let $L \in SLT$.  Then, there exist $k, w \in \sg^k$ such that $w$
is a constant for $L$. That is, $xwy,pwq \in L$ implies $xwq,pwy~
\in L$. Consider $xy^lz \in L.  Then, xyy^kz, xy^kyz  \in L$ implies
that $xy^{k+2}z \in L$.  Hence $xy^{l+1}z  \in L$ which implies $L
\in ESF$. Hence $SLT \in ESF$. But the converse is not true. The
language $ \{ abb^+c, pbb^+q\}$ is ESF, but for no $k \geq 1$ ,
$SLT$ property holds.
\end{proof}

\begin{theorem}
The relations in figure \ref{rel} hold; The arrows indicate strict
inclusions and every two families not linked by a path in this
diagram are incomparable.
\end{theorem}
\begin{proof}
This diagram appears in \cite{MPS} with out the GSCO classes.  Hence
all relations
 between families other than GSCO classes are known.
\begin{enumerate}
\item $COMB \subset TSyGSCO$. Let $L\in COMB$, i.e. $L=\sg^*U,~U\subseteq \sg$. $\sg^*U$
is closed with respect to the GSCO operation. This implies $\sg^*U$
is a crossover language. Hence $L\in TSyGSCO$.
\par This inclusion is strict. $a^*b^*\in TSyGSCO$ but $\notin COMB$.
\item DEF and TSyGSCO are incomparable.
\par $ab^+\in TSyGSCO-DEF$. $(a+b)^+aabb\in DEF-TSyGSCO$. Since $(a+b)^+aabb$ is
not a crossover language.
\item TSyGSCO and NIL are incomparable.
\par $\{a^2,a^3\}\in NIL-TSyGSCO$. $a^*b^*\in TSyGSCO-NIL$.
\item TSyGSCO and COMM are incomparable.
\par $\{ab,ba\}\in COMM-TSyGSCO$. The other way is obvious.
\end{enumerate}

\end{proof}

\section{Conclusion}
We have presented a new operation $GSCO$ over words and languages,
which in some sense abstracts the cross-over of chromosomes in the
living organisms. This study of $ GSCO$ reveals many interesting
results, such as $GSCO^*(L)$  is regular for any $L$. This result
could be useful in places where a generation of regular languages
are required. 
\par we conclude this paper by pointing out some further directions of research.
A study of generalised parallel cross over of words and languages, where the
parallelism is allowed, (i.e. cross over may occur more than one places) can be
initiated and a comparison between the generalised sequential crossover and
generalised parallel crossover has the potential of bringing results of worth.
\par Though this study has come out with a characterisation of strictly locally
testable languages (SLT) in terms of $GSCO$, this result does not
compare the characterisations of SLT, which are available earlier
with the newly obtained one. That is, the characterisations of SLT
could be compared in the sense of complexity, which is worth
investigating.
\par In our opinion the construction of $B$ set can  be used for data compression
in the following sense. To store a
 crossover language $L$, which is closed under $GSCO$, it is sufficient to store the
sets $B,~S,~E$ . $L$ can be retrieved from these by iterated GSCO
operation.



\end{document}